\documentclass[11pt]{article}
\usepackage[margin=1in]{geometry}
\usepackage[T1]{fontenc}
\usepackage[utf8]{inputenc}
\usepackage{lmodern}
\usepackage{microtype}
\usepackage{amsmath,amssymb,amsthm}
\usepackage{graphicx}
\usepackage{xcolor}
\usepackage{url}
\usepackage[colorlinks=true,linkcolor=blue,citecolor=blue,urlcolor=blue]{hyperref}

\graphicspath{{./}}

\newcommand{\multirow}[3][]{#3}
\newcommand{\Tr}{\mathrm{Tr}}
\newcommand{\sfX}{\mathsf{X}}
\newcommand{\sfZ}{\mathsf{Z}}
\newcommand{\sfW}{\mathsf{W}}
\newcommand{\cH}{\mathcal{H}}

\newcommand{\ket}[1]{\lvert #1\rangle}

\newtheorem{thm}{Theorem}[section]
\newtheorem{lem}[thm]{Lemma}
\newtheorem{cor}[thm]{Corollary}
\theoremstyle{definition}
\newtheorem{defn}[thm]{Definition}
\theoremstyle{remark}
\newtheorem{remark}[thm]{Remark}

\newenvironment{proofa}[1][Proof]{\par\noindent\textit{#1. }\ignorespaces}{\par}
\newenvironment{proofof}[1]{\begin{proof}[Proof of #1]}{\end{proof}}

\title{Verifier-initiated quantum message-authentication via quantum zero-knowledge proofs}
\author{Wusheng Wang\thanks{Graduate School of Mathematics, Nagoya University, Chikusa-ku, Nagoya 464--8602, Japan. Email: \texttt{wang.wusheng.g1@s.mail.nagoya-u.ac.jp}} \and Masahito Hayashi\thanks{School of Data Science, The Chinese University of Hong Kong, Shenzhen, Shenzhen 518172, China; International Quantum Academy, Shenzhen 518048, China; and Graduate School of Mathematics, Nagoya University, Nagoya 464--8602, Japan. Email: \texttt{hmasahito@cuhk.edu.cn}}}
\date{}

\begin{document}
\maketitle
\begin{abstract}
On-demand authentication is critical for scalable quantum systems, yet many existing quantum signature and message-authentication schemes are signer-initiated, requiring advance distribution of authentication material even when no verification occurs. We introduce verifier-initiated quantum digital signatures (VIQDS), in which the verifier requests authentication only when needed and the signer responds once; after issuance, verification proceeds without further interaction. Practically, shifting authentication to a verifier-driven, on-demand workflow reduces avoidable communication and storage overhead and aligns with deployments where verification is sporadic, such as distributed services and audit-oriented infrastructures. Our approach leverages quantum zero-knowledge techniques so that verification reveals nothing about the signer's secret key beyond the fact that the signature is valid. We present a general conversion principle from suitable quantum proof protocols to VIQDS, together with a concrete realization based on elementary qubit platforms. Here, we show information-theoretic security against forgery and privacy against curious verifiers without computational hardness assumptions.
\end{abstract}

\section{Introduction}
Digital signatures are a cornerstone of modern cybersecurity, providing guarantees of integrity, authenticity, and non-repudiation across scientific, governmental, and industrial systems. As quantum technologies emerge, researchers have proposed quantum versions of authentication and signing. 
Recent experimental progress has shown a strong feasibility of real quantum digital signature applications: H. Yin et al. demonstrated a quantum digital signature for signing a document with at most $2^{64}$-length\cite{yin2023}. Chapman et al. implemented an entanglement-based quantum digital signature over campus network with sufficiently low qubit error rates($<5\%$)\cite{chapman2024}. J. Xu et al. obtained a semiconductor-based quantum network combining quantum Byzantine agreement and quantum digital signature\cite{jing2024}. Y. Du et al. achieved a chip-integrated quantum signature network over 200 km\cite{du2025}. X. Cao et al. demonstrated a quantum e-commerce scheme based on quantum digital signature that achieves a signing rate of $0.82$ times per second for an agreement size of $0.428$ megabit\cite{cao2024}.
However, most existing quantum designs assume a \emph{signer-initiated} model, where the signer proactively distributes authentication data before it is needed. In many real-world scenarios—such as decentralized networks or large-scale administrative workflows—this approach is inefficient. To address this, we introduce the concept of a \emph{verifier-initiated quantum digital signature} (VIQDS), a protocol where the verifier requests authentication only when needed. This on-demand pattern reduces unnecessary pre-distribution and aligns with scalable verification in blockchain systems and other high-throughput environments \cite{Fang2020,Kumara23,NIST80063A}.
Classical cryptography has long shown that a \emph{verifier-initiated} approach to signing and authentication can be implemented efficiently using \emph{zero-knowledge proofs} (ZKPs). A zero-knowledge proof is a protocol that allows a prover to convince a verifier that a statement is true without revealing any additional information beyond its validity \cite{Goldwasser85,Fiat86,Guillou88,goldreich1991,Okamoto92,Katz20}. Building on these successes, we argue that the same paradigm is equally important in quantum settings. In this work, we extend the concept to \emph{quantum zero-knowledge proofs} (QZKPs), which serve as the natural foundation for constructing secure verifier-initiated protocols in the quantum domain.

From a security perspective, the classical reliance of expressive zero-knowledge proofs (ZKPs) on computational assumptions raises serious concerns in the presence of quantum adversaries. Foundational results show that one-way functions are necessary and sufficient for computational zero knowledge for all of NP \cite{goldreich1991,Katz20,hirahara2024}; yet many classical one-way functions (e.g., factoring-based) become vulnerable under quantum algorithms. This tension has led to two complementary directions. The \emph{post-quantum} approach seeks to preserve classical ZKP constructions by replacing fragile assumptions with ones believed to be quantum-resistant, such as transparent and hash-based designs or lattice-based hardness assumptions \cite{esgin2019,esgin2023,ishai2021,kales2022}, often using techniques like re-randomization for module-LWE to enable efficient proofs \cite{steinfeld2022}. In contrast, the \emph{quantum-native} approach abandons computational assumptions entirely and employs quantum zero-knowledge proofs (QZKPs) that, in some cases, achieve information-theoretic security grounded in physical principles \cite{kobayashi2008}. This makes quantum-native methods particularly appealing for long-term security, as guarantees based on the laws of physics remain robust even against future breakthroughs in cryptanalysis.

Despite these advances, a critical gap remains: no existing work provides a fully specified \emph{VIQDS} protocol
while existing studies \cite{Zawadzki2018,Chen2023} discussed
verifier-initiated quantum identification protocols.
Current QZKP-based authentication frameworks—including those formalized in \cite[Definition 4.2]{morimae2022}, \cite[Definition 4.1]{MY22}, and \cite[Definition 9]{CM24}—implicitly assume \emph{signer-initiated} workflows. While suitable for certain contexts, signer initiation imposes persistent communication and computational overhead, as the signer must proactively 
perform authentication steps for documents that may never be examined. 
By contrast, a verifier-initiated approach—while conceptually beneficial even in quantum settings—has so far only been proposed in classical cryptographic systems, where it enables on-demand retrieval, reducing unnecessary pre-distribution and improving scalability in high-throughput or decentralized environments \cite{Fang2020,Kumara23,NIST80063A}.
Closing this gap by establishing a practical and secure VIQDS protocol represents both a conceptual advance—moving beyond classical paradigms—and a practical step toward efficient, post-quantum authentication for future quantum infrastructures (Table \ref{fig1}).

Table~\ref{fig1} highlights three limitations of existing approaches.
First, most quantum message-authentication/signature schemes are signer-initiated, which incurs avoidable pre-distribution overhead when verification is on-demand.
Second, existing verifier-initiated quantum proposals mainly focus on identity authentication rather than message authentication.
Third, current QZKP-based authentication frameworks for messages implicitly follow signer-initiated workflows, leaving the verifier-initiated message-authentication setting unaddressed.

A second, more subtle gap concerns how QZKP security is modeled when used as the core of an authentication protocol. The \emph{quantum interactive proof} (QIP) framework and its complexity-theoretic foundations provide powerful abstractions for properties such as completeness and soundness \cite{watrous1999,jain2011}. Building on these foundations, several QZKP paradigms have emerged: schemes based on the indistinguishability of quantum states \cite{watrous2002}, hybrid approaches combining Hamiltonian measurements with zero-knowledge techniques for QMA \cite{vidick2020,mahadev2018}, and constructions secured through certified deletion \cite{broadbent2016,hiroka2022,broadbent2020}, alongside seminal complexity results for quantum zero knowledge \cite{watrous2006}. However, these models often overlook practical adversarial behaviors relevant to authentication, such as dishonest provers or curious verifiers, leaving a critical gap between theoretical guarantees and real-world security requirements.

Prevailing soundness definitions often fail to capture \emph{dishonest provers} who deviate from prescribed interactions in ways unrelated to any valid witness. In authentication, such deviations create concrete forgery opportunities if left unmodeled. On the verifier side, the traditional dichotomy between {\it honest} and {\it fully dishonest} is also too coarse. Inspired by the notion of a \emph{specious} adversary in quantum private information retrieval—an indistinguishable-but-curious entity—we argue that QZKPs must address verifiers who attempt to extract secret information while maintaining transcripts that appear honest \cite{Baumeler15,Aharonov19,song2024}. Because an honest prover cannot distinguish a specious verifier from a truly honest one, zero-knowledge guarantees must be robust against this subtle yet realistic threat.

As our goals and contributions, we address both gaps identified above to enable verifier initiation for quantum-secure signatures and ensure practical implementability.
First, we refine the QZKP framework to explicitly incorporate adversarial models, accounting for both \emph{dishonest provers} and \emph{specious verifiers}, thereby ensuring robustness against these realistic threats.
Second, we develop a general concatenation method for the refined QZKP framework by which the soundness can be exponentially enhanced.
Third, we introduce a \emph{general compiler} that systematically transforms any QZKP satisfying these properties into a \emph{verifier-initiated quantum digital signature (VIQDS)} protocol. Under this transformation, completeness of the underlying QZKP guarantees completeness of VIQDS, while security against dishonest provers and specious verifiers is preserved.
Fourth, we present a concrete, highly efficient QZKP protocol that leverages the interplay between measurement-induced disturbance and the eigenstructure of observables.
Importantly, this protocol is not only theoretically sound but also experimentally feasible: it can be implemented using current qubit technologies such as photonic or trapped-ion platforms, without requiring exotic hardware or unproven primitives.
Integrating this QZKP into our generic conversion yields a practical VIQDS protocol with the desired interaction pattern and end-to-end security.
The relationship between VIQDS and QZKP is illustrated in Figure~\ref{F3}.

In this work, we first formalize the quantum interactive proof setting and the associated notions of completeness, soundness, and quantum/specious zero knowledge \cite{watrous1999,jain2011,watrous2002,watrous2006,vidick2020,mahadev2018,broadbent2016,hiroka2022,broadbent2020}, and then present a general transformation from QZKP to VIQDS together with the classical motivations for verifier initiation \cite{Goldwasser85,Fiat86,Guillou88,goldreich1991,Okamoto92,Katz20}
at Results. 
Then, we introduce the discrete Heisenberg preliminaries, specifies the concrete protocol, and proves completeness, soundness, and specious zero knowledge, including privacy against classical specious verifiers (and hence also quantum specious verifiers), and also discuss the experimental feasibility of the protocol under current technologies
at Methods.

\section{Results}
\subsection{Formulation of VIQDS}
In classical cryptography, verifier-initiated authentication has long been realized through zero-knowledge proofs (ZKPs).  
Examples include Fiat-Shamir, Schnorr, and zk-SNARKs.  
In these protocols, the verifier requests a proof while 
the prover responds only when prompted.  
Such an on-demand model is efficient and scalable, which 
is widely used in systems like secure login and anonymous credentials.

In contrast, existing quantum digital signature schemes — including those based on quantum zero-knowledge proofs \cite{morimae2022, MY22, CM24} — mostly follow the \emph{signer-initiated} model  
where the signer generates and distributes signatures in advance.  
This is inefficient in large-scale or decentralized systems
that take on-demand verification as the norm.

To close this gap, we introduce the \emph{verifier-initiated quantum digital signature} (VIQDS).  
Here, the verifier — not the signer — initiates authentication.  
The signer generates a signature only when requested.  
This avoids precomputation and pre-distribution. In addition, it
has a lot of merits, including lower communication overhead, stronger scalability, and resource-conscious authentication.

The VIQDS protocol comprises one $\emph{signer}$ (Alice) and $N$ $\emph{participants}$. One participant, $\emph{Bob}$, is designated as the $\emph{verifier}$. A key assumption is that each participant $i$ possesses an ideal quantum memory, denoted by $\mathcal M_i$. The protocol is structured into four distinct phases:

\begin{itemize}
    \item In the {\it key-generation} phase, 
    Alice generates a private key $sk$ and prepares $N$ identical quantum public key states $pk$. She distributes one copy to each participant $i$, who stores it in $\mathcal M_i$. This establishes the trust infrastructure without message-specific computation (Figure \ref{FX2}a).
    
    \item In the {\it pre-signing} phase, when Bob wants to verify a message $m$, he sends a quantum challenge — a quantum state $\tilde{\rho}$ — to Alice.  
    This is a request for authentication.  
    It ensures the signature is generated only on-demand, as shown in Figure \ref{FX2}b.
    
    \item In the {\it signing} phase, Alice receives the challenge and $m$.  
    She generates a signature $sgn$ using $sk$.  
    She sends $(sgn, m)$ to Bob.  
    The signature is created only when requested, as shown in Figure \ref{FX3}a.
    
    \item In the {\it verification} phase, Bob verifies the signature using $pk$, $m$, and $sgn$.  
    He outputs 1 (accept) if valid, 0 (reject) otherwise.  
    The verification is non-interactive.  
    It completes without further communication, as shown in Figure \ref{FX3}b.
\end{itemize}

The above structure gives the verifier control over when and for which message authentication occurs, 
which is critical for real-world systems where efficiency and scalability matter.
In the one-copy setting considered here, each participant, including the verifier, stores exactly one copy of the quantum public-key state $pk$.

Our one-copy, one-time setting is defined by public-key consumption and refresh as follows.
In our security model, each participant (including the verifier) stores exactly one copy of the quantum public key state $pk$.
Verification is specified as a binary POVM acting on $(pk,m,\mathrm{sgn})$, and therefore the stored $pk$ is treated as a one-time token:
a single authentication/verification attempt consumes the public key and we do not assume reusability.
Hence, one stored key copy supports exactly one authentication.
To authenticate further messages, the signer refreshes the public key by rerunning the key-generation phase and redistributing a fresh $pk$ state when needed.

This one-copy regime is deliberate: it captures on-demand authentication settings where public-key states are distributed as physical tokens and are consumed by verification measurements. In this sense, VIQDS can be viewed as a publicly verifiable signing/authentication primitive with explicit key rotation, akin to one-time signature usage where each verification consumes the corresponding public-key token. Our security definitions and proofs are stated for this regime; extending to multi-copy public keys requires separate modeling and analysis.

The reason that a one-copy quantum public key is essential in our setting is as follows.
The role of the one-copy assumption is best understood by contrasting the quantum and classical settings.
If the public key were a \emph{classical} object (a fully readable classical description), then the verifier could directly obtain the identification information of the public key from that description and mount the corresponding attack; in this sense, the one-copy restriction would be vacuous in the classical setting.
In our construction, however, the public key is an \emph{unknown quantum state} held in a single copy.
The most direct attack route is to identify the public-key state and use the resulting classical description to mount an attack, but such identification cannot be achieved from a single copy: full state reconstruction (tomography) fundamentally requires many identical copies.
Therefore, restricting the verifier to one copy rules out this direct identification-based route, which is precisely where the quantum nature of $pk$ becomes essential in our verifier-initiated setting.

Next, we discuss the security against curious but transcript-indistinguishable adversaries.
Each participant receives a copy of the public key during key generation.  
They may try to extract secret information while appearing honest.  
To capture this threat, we introduce the notion of a \emph{specious participant}.  
Speaking informally, a \emph{specious participant} behaves indistinguishably from an honest participant in observable transcripts  
while they may try to learn the signer's private key.
We define two types of security:
\begin{itemize}
    \item 
    The security against a {\it quantum specious participant} is defined as follows.
    When acting as the \emph{verifier}, this participant interacts with the signer in a way that does not change the signer's quantum state at the signing phase.  
    Specifically, suppose the participant sends a quantum challenge to the signer, just like an honest verifier would.  
    Then, the quantum state in the signer's memory at the signing phase is exactly the same as it would be if the participant were honest.  
    From the signer's perspective, there is no detectable difference.  
    Yet, the participant may be secretly measuring or manipulating their own copy of the public key to extract information.
    
    \item 
        The security against a {\it classical specious participant} is defined as follows.
When acting as the \emph{verifier}, this participant interacts with the signer in a way that does not change the signer’s classical memory at the signing phase.  
    Specifically, suppose the participant sends a quantum challenge to the signer, just like an honest verifier would.  
    Then, the classical variables stored in the signer’s memory at the signing phase follow the same statistical distribution as they would if the participant were honest.  
    Again, the signer cannot tell the difference.  
    But the participant may be trying to learn the private key through side channels or other means.
\end{itemize}
When a participant plays the role of the verifier in VIQDS, we refer to it as a (quantum/classical) specious verifier in the underlying VIS/QZKP formulation.
These definitions reflect realistic adversarial behaviors.  
The adversary does not break the protocol’s observable behavior.  
However, they try to extract secret information.  
Security against such adversaries ensures that the signer’s private key remains protected — even if a participant is curious or malicious.

The reasons why specious participants are a realistic threat are as follows.
In many real-world systems, participants are not assumed to be fully honest.  
On the other hand, they are also not assumed to be fully malicious.  
Instead, they may behave \emph{speciously} — appearing honest while secretly trying to gain an advantage.
This behavior is plausible for two reasons:

\begin{itemize}
    \item 
    If a participant's dishonest behavior is not detected, they face no penalty.  
    This creates an incentive to try to extract information without being noticed.
    
    \item 
    If their specious behavior is detected — for example, if they cause a verification failure or leave a detectable trace — they may face penalties such as being excluded from the system or losing reputation.  
    However, if they remain undetected, they gain secret information without cost.
\end{itemize}

This is why we model specious participants.  
They represent a subtle but realistic threat: adversaries who are rational and risk-averse.  
They will not break the protocol in an obvious way.  
But they will try to exploit it if they can do so without being caught.

To rigorously evaluate VIQDS, we introduce the following key notions as formal security properties. 
A protocol is said to satisfy $\alpha$-{\it completeness} if, whenever Alice and Bob act honestly, Bob accepts the signature with probability at least $\alpha$.
A {\it forging attack} is defined as follows. An adversary intercepts communication during the pre-signing and signing phases. The verifier requests a signature on $m$. The adversary modifies the pre-signing phase so that the signer is asked to sign $m' \neq m$, and then modifies the signing phase so that the verifier accepts the signature for $m'$.
The protocol is said to be information-theoretically unforgeable with quantum specious participants if any adversary succeeds in a forging attack with probability at most $\epsilon$ when all colluding participants are quantum specious and both the signer and the verifier are honest.
Similarly, the protocol is said to be information-theoretically unforgeable with classical specious participants if any adversary succeeds in a forging attack with probability at most $\epsilon$ when all colluding participants are classical specious and both the signer and the verifier are honest.

The reason why this security model matters, and how it mitigates dishonest behavior, can be explained as follows.
Traditional models typically treat adversaries as either fully honest or fully malicious.
Our notion of \emph{(quantum) specious} participants captures an intermediate and realistic threat:
an adversary may actively interact with the signer while keeping the interaction
\emph{indistinguishable from honest behavior from the signer’s perspective}.
Concretely, a specious adversary is constrained so that the signer’s quantum state (and, in particular,
its classical memory distribution) at the signing phase remains the same as in an honest execution,
even though the adversary may measure or manipulate \emph{its own} public-key copy to extract information.
By proving security against such adversaries---namely, that any forging attack succeeds with probability at most
$\epsilon$---we obtain robustness beyond conventional honest/dishonest dichotomies, supporting practical deployment
in settings where participants may be curious yet risk-averse.

A further concern in multi-participant deployments is the \emph{unauthorized transfer} of a public-key copy to another party
who later acts as a verifier. Accumulating additional copies could, in principle, increase the recipient’s ability to
extract information and thereby raise the risk of key compromise or forgery attempts.
To discourage such collusion, we allow an optional auditing mechanism in which the signer (as the prover in the underlying
VIS/QZKP view) can randomly designate certain participants to act as verifiers.
While this option is not essential to the core verifier-initiated workflow, it provides a practical deterrent against
public-key sharing in distributed environments.

Importantly, our VIQDS formulation itself is \emph{primitive-agnostic}:
it is specified purely in terms of roles, phases, and security goals.
We next connect this formulation to quantum zero-knowledge proofs via VIS protocols, which yields
a concrete and secure VIQDS construction.

\subsection{QIP: Foundations for Zero-Knowledge in VIQDS}
The core idea behind VIQDS is to repurpose the roles in a quantum zero-knowledge proof (QZKP):  
the prover becomes the signer,  
and the verifier takes the role of the verifier in VIQDS.
To make this conversion secure and meaningful,  
we first need to establish a quantum interactive proof (QIP) framework  
that reflects real-world adversarial behavior —  
not just honest or passive parties,  
but also dishonest provers and curious verifiers.
This is essential for VIQDS.  
In applications like blockchain or decentralized systems,  
the verifier initiates authentication on-demand.  
It must be protected against both forged signatures and information leakage.

We now describe the basic structure of a quantum interactive proof game.  
It involves two parties: a prover (P) and a verifier (V).  
The verifier wants the prover to prove a statement $r$.  
If the proof succeeds, the verifier outputs 1; otherwise, 0.  
The game proceeds over $m$ rounds.  
Both parties hold quantum registers: P has register $P_j$, and V has register $V_j$.
In round 1, the verifier prepares a quantum state using 
the quantum operation $Z_1$, and sends a quantum message $X_0$ to the prover.  
In round 2, the prover applies the quantum operation $Z_2$ to its register, and sends back a message $Y_0$ (quantum or classical).  
In round 3, the verifier applies the quantum operation $Z_3$, sends message $X_1$, and so on.  
After $m-1$ rounds, the verifier applies the final quantum operation $Z_{m}$ and outputs a binary decision — 1 for acceptance, 0 for rejection.
Here, each quantum operation $Z_{j}$ is given as a trace-preserving completely positive (TP-CP) map.

We now introduce the verifier-initiated $\Sigma$-like protocol (VIS protocol). A schematic of the underlying 3-round QIP game is illustrated in Figure~\ref{F1}. We adopt this structure to emphasize two key aspects: it preserves the three-round format characteristic of classical $\Sigma$-protocols, yet differs fundamentally in that the interaction is initiated by the verifier rather than the prover.
In the classical setting \cite{stanfordCS355,zhang2023sigma,ietfSigmaProtocol}, a $\Sigma$-protocol begins with a commitment from the prover, 
whereas our protocol starts with a challenge from the verifier.

Here, the {\it statement} is denoted by $r$, which the prover aims to prove knowledge of. The public information consists of a set of quantum systems $\{\mathcal{H}_r\}_r$, prepared in advance and shared with the verifier, and stored in the verifier's quantum memory $\mathcal{M}$. In VIQDS, these systems become the {\it public key}. 
The prover's private information is a witness $W_r$ for each statement $r$. In VIQDS, this witness becomes the signer's {\it private key}. Importantly, for $r \neq r'$, the witnesses $W_r$ and $W_{r'}$ are independent, ensuring that compromising one signature does not affect others.

VIS protocol proceeds in three distinct steps:

\begin{itemize}
\item In Step 1, the verifier sets the register $V_0$ to be 
the system ${\cal H}_r$, 
and prepares a quantum challenge using 
the quantum operation $Z_1$, and sends 
a quantum challenge $X_0$ 
to the prover. 
This corresponds to the verifier's request for authentication in VIQDS.
\item In Step 2, the prover sets the register $P_0$ to be $W_r$.
Upon receiving the challenge $X_0$, 
the prover applies the quantum operation $Z_2$ to 
the composite system $P_0 \otimes X_0$.
It then generates a response $Y_0$ and sends it back to the verifier. 
This corresponds to the signer’s generation of a signature in VIQDS.
\item In Step 3, the verifier applies the quantum operation $Z_3$ to verify the prover’s response. 
The outcome $X_1$ is binary, 1 (accept) or 0 (reject). 
This corresponds to the verifier’s verification of the signature in VIQDS.
\end{itemize}

Hence, a VIS protocol is given as a tuple 
($\{\mathcal{H}_r\}_r$, $\{W_r\}_r$, 
$V_0$, $V_1$, $P_0$, 
$X_0$, $X_1$, $Y_0$, 
$Z_1$, $Z_2$, $Z_3$).
In this formulation,
$V_0$, $V_1$, $P_0$,
$X_0$, $X_1$, and $Y_0$ are treated as quantum systems,
although some of these systems may be initialized in classical states.

Once two VIS protocols are given,
we can formulate the {\it concatenation} of these two protocol as follows.
Concatenation is useful for considering the scalability.
Consider two VIS protocols characterized by
${\cal V}^1=$
($\{\mathcal{H}_r^1\}_r$, $\{W_r^1\}_r$, 
$V_0^1$, $V_1^1$, $P_0^1$, 
$X_0^1$, $X_1^1$, $Y_0^1$, $Z_1^1$, $Z_2^1$, $Z_3^1$)
and
${\cal V}^2=$
($\{\mathcal{H}_r^2\}_r$, $\{W_r^2\}_r$, 
$V_0^2$, $V_1^2$, $P_0^2$, 
$X_0^2$, $X_1^2$, $Y_0^2$, $Z_1^2$, $Z_2^2$, $Z_3^2$).
Then, we define their concatenation 
${\cal V}^1\otimes {\cal V}^2$ 
with the tuple
($\{\mathcal{H}_r^{1,2}\}_r$, $\{W_r^{1,2}\}_r$, 
$V_0^{1,2}$, $V_1^{1,2}$, $P_0^{1,2}$, 
$X_0^{1,2}$, $X_1^{1,2}$, $Y_0^{1,2}$, 
$Z_1^{1,2}$, 
$Z_2^{1,2}$, $Z_3^{1,2}$).
Here, we define
($\{\mathcal{H}_r^{1,2}:=\mathcal{H}_r^1\otimes \mathcal{H}_r^2$, 
$W_r^{1,2}:=
(W_r^1,W_r^2)\}_r$, 
$V_0^{1,2}:=V_0^1\otimes V_0^2$, 
$V_1^{1,2}:=V_1^1\otimes V_1^2$, 
$P_0^{1,2}:=P_0^1\otimes P_0^2$, 
$X_0^{1,2}:=X_0^1 \otimes X_0^2$, 
$Y_0^{1,2}:=Y_0^1\otimes Y_0^2$, 
$Z_1^{1,2}:=Z_1^1\otimes Z_1^2$, 
$Z_2^{1,2}:=Z_2^1\otimes Z_2^2$.
Only the definitions of $Z_3^{1,2}$ and $X_1^{1,2}$ are slightly different as follows.
Applying $Z_3^1\otimes Z_3^2$, the verifier obtains two bits $X_1^1 $ and $X_1^2$.
Then, the verifier outputs 1 only when both bits are 1, i.e., 
we define $X_1^{1,2}:=X_1^1 X_1^2$.
Here, the states on 
${\cal H}_r^1$, ${\cal H}_r^2$ for $r$
are independently prepared.
In particular, when a protocol $\mathcal{V}$ is concatenated $l$ times,
 the combined protocol is written as $\mathcal{V}^{\otimes l}$.

We now discuss the {\it security} of this 3-round structure. This structure is not arbitrary; rather, it strikes an effective balance between efficiency and security:
The quantum challenge in round 0 prevents precomputation. 
The classical response in round 1 enables efficient verification. 
The structure naturally supports the verifier-initiated flow: the verifier requests a signature on-demand, and the signer responds only when needed.
To ensure security, we require two fundamental properties:  
completeness and soundness.

{\it Completeness} means:  
If the verifier follows the protocol,  
and the prover follows the protocol with the correct input $(r, W_r)$,  
then the verifier accepts with probability at least $\alpha$.  
This guarantees that an honest signer can always authenticate a message —  
a basic requirement for any digital signature.
{\it Soundness} must account for different types of adversarial behavior.  
We consider two distinct cases:

\begin{itemize}
    \item Soundness of {\it type 1} is defined as follows.
    The prover is honest but mistaken — for example, using a wrong private key.  
    Formally, if the verifier follows the protocol and the prover uses an incorrect witness $(r, W_r')$ where $W_r' \neq W_r$,  
    the verifier accepts with probability at most $\beta$.
    
    \item Soundness of {\it type 2} is defined as follows.
    The prover is dishonest and tries to forge a signature without any valid private key.  
    Formally, if the verifier follows the protocol and the prover uses a protocol that depends on $r$ but is independent of any valid witness $W_r$,  
    the verifier accepts with probability at most $\beta$.
\end{itemize}

These two forms of soundness are not redundant.  
They address fundamentally different threats:  
one is an honest-but-mistaken signer,  
the other is an actively malicious forger.  
Both are common in real-world authentication systems.
We say a quantum interactive proof system satisfies $\beta$-soundness if it satisfies both type 1 and type 2.  
This dual soundness is essential for VIQDS.
Formally, we denote an quantum interactive proof game as $QIP_{\alpha\beta}$ if it has $\alpha$-completeness and $\beta$-soundness.

However, in authentication applications, completeness and soundness alone are not enough.  
The proof relies on secret knowledge (the witness).  
If the verifier is dishonest or curious, he may try to learn this secret.  
To prevent this risk, we must add the property of {\it zero-knowledge}.  
This is particularly important for VIQDS:  
the verifier should learn nothing about the signer’s private key —  
even if he is curious or specious.

\subsection{QZKP: Protecting against specious verifiers in VIQDS}
Building on the quantum interactive proof (QIP) framework established in the previous section, 
we now focus on a critical extension: ensuring that the verifier learns nothing about the prover’s secret — 
even when the verifier behaves speciously. 
While QIP provides the structural foundation for interaction and security properties such as completeness and soundness, 
it does not inherently guarantee privacy of the witness. 
To address this, we introduce the concept of quantum zero-knowledge proofs (QZKPs), 
which augment QIP protocols with privacy guarantees essential for secure authentication in VIQDS.

In particular, we refine the adversarial model introduced in QIP — including specious verifiers — 
and define corresponding zero-knowledge properties that ensure the signer’s private key remains protected 
even under subtle and realistic threats. 
This refinement leads naturally to the concept of \emph{quantum zero-knowledge proofs} (QZKPs), 
which extend QIP by guaranteeing that the verifier learns nothing beyond the validity of the statement.

To capture realistic adversarial behavior in quantum authentication, we introduce three types of verifiers — honest, specious, and dishonest — and define corresponding zero-knowledge properties.
A verifier is said to be {\it honest} if it always follows the protocol exactly. In VIQDS, this corresponds to a standard non-adversarial verifier.

A verifier is said to be {\it quantum specious} if, at each step, it interacts with the prover in a way that leaves the prover's quantum state unchanged. More precisely, when the prover is honest, the joint quantum state on the prover's local system, including its classical memory, is identical at each verifier step $i$ to what it would be under an honest verifier. In VIQDS, this models a verifier that attempts to learn the private key without altering the observable transcript.

A verifier is said to be {\it classical specious} if, at each prover step, it interacts with the prover in a way that leaves the prover's classical memory unchanged. More precisely, when the prover is honest, the classical variables stored in the prover's memory follow the same statistical distribution at each prover step $i$ as they would under an honest verifier. In VIQDS, this models a verifier that may attempt to learn the witness through classical side channels without changing the prover's classical view of the interaction.

Note that any quantum specious verifier is also a classical specious verifier.  
Importantly, because the messages sent by a specious verifier are identical to those sent by an honest verifier, the prover cannot distinguish between them — making this a subtle but realistic threat in quantum authentication.
We call a verifier {\it dishonest} if it may perform arbitrary deviations from the protocol, for example by measuring quantum states in unintended bases or by deviating from the protocol entirely. In VIQDS, this corresponds to a fully adversarial verifier.

Based on these verifier types, we introduce four levels of quantum zero-knowledge. A protocol is said to be {\it honest-verifier QZKP} (HVQZKP) if, when both parties are honest, the verifier's final state is independent of the witness $W_t$. This is the weakest form and is suitable only for non-adversarial settings.
A protocol is said to be {\it quantum-specious-verifier QZKP} (QSVQZKP) if, when the prover is honest and the verifier is quantum specious, the verifier's state at each step is independent of the witness $W_t$. This notion is crucial for VIQDS, because it ensures that even a curious but transcript-indistinguishable verifier cannot learn the private key.
A protocol is said to be {\it classical-specious-verifier QZKP} (CSVQZKP) if, when the prover is honest and the verifier is classical specious, the verifier's state at each step is independent of the witness $W_t$. Here, {\it classical specious} means that at each prover step the prover's classical memory has the same distribution as in the honest-verifier interaction, while no constraint is imposed on the prover's quantum state.
A protocol is said to be {\it dishonest-verifier QZKP} (DVQZKP) if, for a dishonest verifier, the verifier's state at each step depends only on public information and not on the witness. This is the strongest form and is suitable for fully adversarial settings.

These notions induce two related hierarchies, and it is important to distinguish them.
At the level of verifier models, a quantum-specious verifier is, by definition, also classical-specious
(i.e., it satisfies the classical-specious requirement as well).
However, when we speak of $QZKP$---namely, protocol classes defined by zero-knowledge against a given verifier model---the inclusion goes in the opposite direction:
zero-knowledge against classical-specious verifiers implies zero-knowledge against quantum-specious verifiers.
Accordingly, among protocol classes, $HVQZKP$ is the weakest and $DVQZKP$ is the strongest, and in particular $CSVQZKP \subseteq QSVQZKP$.
Thus, in our one-copy setting, $CSV$-verifier zero-knowledge implies $QSV$-verifier zero-knowledge.

Figure~\ref{F2} illustrates the relationships between these forms.
Similar to $QIP_{\alpha,\beta}$, we write $QZKP^{t}_{\alpha,\beta}$ for a quantum zero-knowledge proof with $\alpha$-completeness and $\beta$-soundness, where $t \in \{\mathsf{HV}, \mathsf{QSV}, \mathsf{CSV}, \mathsf{DV}\}$ denotes the verifier type.

\begin{thm}\label{TB1}
When the original VIS protocols
${\cal V}^1$ and ${\cal V}^2$ 
are 
$QZKP^{t}_{\alpha_1,\beta_1}$
and
$QZKP^{t}_{\alpha_2,\beta_2}$, respectively with
$t \in \{\mathsf{HV}, \mathsf{QSV}, \mathsf{CSV}, \mathsf{DV}\}$,
then the concatenated VIS protocol 
${\cal V}^1\otimes {\cal V}^2$ 
is $QZKP^{t}_{\alpha_1\alpha_2,\beta_1\beta_2}$.
In particular,
when the original VIS protocol ${\cal V}$ 
is $QZKP^{t}_{\alpha,\beta}$ with 
$t \in \{\mathsf{HV}, \mathsf{QSV}, \mathsf{CSV}, \mathsf{DV}\}$,
then the concatenated VIS protocol 
${\cal V}^{\otimes l}$ is $QZKP^{t}_{\alpha^l,\beta^l}$.
\end{thm}

\subsection{General compiler from VIS protocol to VIQDS protocol}
Our central insight is simple yet powerful:  
a VIS protocol satisfying the condition of QZKP 
can be systematically converted into a verifier-initiated quantum digital signature (VIQDS) protocol via our general compiler.  
Importantly, the resulting VIQDS protocol inherits the security guarantees of the underlying VIS protocol — including completeness, soundness against dishonest signers, and protection against specious verifiers.  
These security properties will be formally proven in Methods.

This conversion works by reassigning roles:  
the prover becomes the signer,  
and the verifier becomes the authenticating party.  
This allows the verifier to initiate authentication on-demand — a critical advantage in practical systems like blockchain, where proactive signature distribution is inefficient.
To make this concrete, we use a VIS protocol as the underlying QZKP.  
It has $\alpha$-completeness and $\beta$-soundness, and can be designed to satisfy specious zero-knowledge.  
The resulting VIQDS inherits these strong security guarantees — robustness against dishonest signers and protection against curious verifiers.
We now present the explicit conversion from the VIS protocol to a VIQDS protocol using our compiler.  
The mapping between elements is summarized in Table~\ref{fig:qzkp_viqds_mapping}.
The protocol proceeds in four phases:

\begin{itemize}
\item In the key generation phase, Alice sets her private key $sk$ as the VIS protocol witness $W_r$ and prepares $N$ identical quantum public key states, $\rho_r$ (denoted as $pk$). She distributes one copy to each participant $i$ for storage in their quantum memory $\mathcal M_i$.
  
    \item In the pre-signing phase,  
    Bob applies circuit $Z_1$, then sends a quantum challenge $X_0$ — along with the message $m$ (treated as statement $r$) — to Alice.  
    This corresponds to the verifier’s Step 0 in the VIS protocol.
    
    \item In the signing phase,  
    Alice applies circuit $Z_2$ using her private key $sk$ and message $m$, then sends a classical response $Y_0$ (the signature) to Bob.  
    This corresponds to the prover’s Step 0 in the VIS protocol.
    
    \item In the verification phase,
    Bob applies circuit $Z_3$ using the public key $pk$, message $m$, and signature $Y_0$, then outputs 1 (accept) or 0 (reject).  
    This corresponds to the verifier's Step 1 in the VIS protocol.
\end{itemize}

For messages of length $\tilde{l}$, the key generation phase requires preparing $L = 2^{\tilde{l}}$ quantum systems as the public key.  
Alternatively, to address scalability for longer messages, we propose a \emph{repeated single-bit scheme}: instead of preparing $L = 2^{\tilde{l}}$ quantum systems for an $\tilde{l}$-bit message, we repeat the $L=2$ case (i.e., one bit) for each of the $\tilde{l}$ bits independently. 
This approach reduces the resource overhead significantly: 
\begin{itemize}
    \item The key generation phase requires only $2\tilde{l}$ quantum systems (instead of $2^{\tilde{l}}$),
    \item The pre-signing phase requires $\tilde{l}$ quantum challenges (one per bit),
    \item 
    The security guarantees that $\epsilon=1/d$ is preserved per bit.  
    In the repeated single-bit scheme, verification is performed independently for each bit instance. The verifier accepts the authenticated payload only if all bit-wise verification tests accept (i.e., the overall rule is an AND over the bit-wise outcomes). For the entire $\tilde{l}$-bit message, we upper-bound the overall forging probability by $\epsilon_{\text{total}} \le \frac{1}{d}$. In the worst case, an adversary may attempt a modification affecting only a single bit instance; hence we use the conservative bound.
\end{itemize}
This method maintains the protocol's structure while enabling efficient scaling for practical message lengths.
As shown in Methods, the security of the VIQDS protocol is guaranteed by the following theorem:

\begin{thm}\label{HJP}
If the original VIS protocol 
is $QZKP^{\mathsf{CSV}(\mathsf{QSV})}_{\alpha,\beta}$,
then the resulting VIQDS protocol satisfies $\alpha$-completeness and
$\epsilon$-information-theoretically unforgeable with classical (quantum) specious participants, where $\epsilon = \beta$.
 \end{thm}

A concrete implementation of this framework, based on the eigenstructure of observables in the discrete Heisenberg representation, is presented in Methods. The $l$-fold concatenation of this protocol operates in dimension $d=p^l$, where $p^l$ is a prime power, and by Theorem~\ref{TB1} it achieves $\alpha$-completeness with $\alpha=1$, that is, perfect success for honest parties, $\beta$-soundness with $\beta=\frac{1}{d}$, that is, forging probability exactly $\frac{1}{d}$, and classical-specious-verifier zero-knowledge.

The above theorem ensures that this concrete QZKP yields a VIQDS protocol with $\alpha=1$-completeness and $\epsilon=\frac{1}{d}$ information-theoretic unforgeability against classical specious participants.
In particular, by setting $p=2$,
the above construction demonstrates that a secure, scalable VIQDS protocol can be realized using existing qubit technologies - with security guarantees that scale inversely with $2^l$.  

Thus, our compiler not only provides a theoretical framework — it enables practical, hardware-compatible quantum authentication with provable, tunable security:  
by increasing the qudit dimension $d = 2^l$, the forging probability $\epsilon = 1/d$ can be made exponentially small,  
while keeping the protocol’s structure — including the number of rounds and quantum systems per message — unchanged.

\section{Discussion}
In this work, we have introduced a high-efficiency quantum zero-knowledge proof (QZKP) protocol that satisfies the standard properties of completeness and zero-knowledge.  
Importantly, it also provides two novel security guarantees:  
- soundness against dishonest provers, and  
- zero-knowledge even in the presence of specious verifiers.

These features address critical gaps in existing QZKP frameworks, which often assume either a passive prover or an honest verifier — assumptions that do not hold in real-world authentication systems.

Our protocol relies on simple, experimentally feasible operations:  
random number generation, eigenstate preparation and measurement, and classical comparison.  
By using an observable’s eigenstates as witnesses, we exploit fundamental quantum mechanics to achieve unconditional security — without relying on unproven computational hardness assumptions.  
This makes our approach particularly valuable for long-term security, where guarantees based on physical principles remain robust against future advances in cryptanalysis.

We have also presented a general compiler:  
any QZKP satisfying our security properties can be systematically transformed into a verifier-initiated quantum digital signature (VIQDS) protocol.  
This compiler allows us to directly translate the strong security of our QZKP — especially its soundness against dishonest provers — into a VIQDS protocol.  
The resulting VIQDS inherits robust unforgeability and protection against specious verifiers.  
These properties are formally proven in Methods, ensuring the protocol’s security is not just intuitive but mathematically grounded.

A key innovation of our work is the use of qubit systems with dimension $d = 2^l$ (a prime power).  
This choice is not merely technical — it is essential for achieving our tight security bound of $\epsilon = 2^{-l}$.  
By leveraging the algebraic structure of finite fields, we ensure that the forging probability is exactly $2^{-l}$, and can be made exponentially small by increasing $l$ — for example, from $l=10$ to $l=20$ — without altering the protocol’s structure or the number of quantum systems per message.

This makes our protocol not only secure, but also scalable:  
even modest qubit systems — realizable with current photonic or trapped-ion platforms — can provide security levels comparable to classical cryptographic schemes, but with unconditional guarantees based on quantum mechanics.

This approach enables practical, scalable quantum authentication — ideal for applications like blockchain, where on-demand verification is essential.
Looking ahead, several directions for future work emerge. These include extending the framework to multipartite settings to enable authentication among multiple parties, which would be important for decentralized systems and multi-signature schemes; optimizing resource overheads, particularly for longer messages, by reducing the number of required quantum systems; and exploring applications to quantum authentication and delegated quantum computation, where verification without revealing secret information is essential.

One limitation of our current protocol is its assumption of noise-free quantum systems and perfect quantum gates.  
In real-world implementations, noise and imperfections may affect performance.  
Future work will investigate error-tolerant variants, potentially using quantum error correction or fault-tolerant techniques to maintain security under realistic conditions.

In conclusion, our results pave the way for robust, practical zero-knowledge proofs in the quantum era.  
By bridging the gap between theoretical security and practical efficiency, we provide a foundation for the next generation of quantum-secure authentication — protocols that are secure, scalable, and aligned with the operational needs of real-world systems.

Finally, although our main focus is the ideal setting, we also provide a robustness analysis against finite-coherence memory, operational imperfections, and detection loss in Supplementary Notes 4 and 5.
There we derive conservative effective completeness/soundness parameters under these non-idealities, together with explicit sufficient feasibility thresholds.
Moreover, as long as the resulting verification gap remains positive, repetition with thresholding continues to yield exponential suppression of honest-reject and forger-accept probabilities (see Supplementary Note 3, Theorem~VI.1).

\section{Methods}

\subsection{Security analysis of the VIQDS protocol}
To show Theorem \ref{HJP}, we now provide the security analysis of the verifier-initiated quantum digital signature (VIQDS) protocol introduced in Results.  
This protocol is constructed by converting a quantum VIS protocol - a type of quantum zero-knowledge proof (QZKP) - using our general compiler.  
The security of the VIQDS protocol directly follows from the security properties of the underlying VIS protocol.
Theorem 1 of Results is shown by the combination of the following theorems.

\begin{thm}
If the original quantum VIS protocol satisfies $\alpha$-completeness, then the resulting VIQDS protocol also satisfies $\alpha$-completeness.
\end{thm}

\begin{proofa}
Suppose the original VIS protocol satisfies $\alpha$-completeness.  
In the VIQDS protocol, Alice (the signer) acts as the prover, and Bob (the verifier) acts as the verifier.  
When both parties follow the protocol — using the correct private key $sk$ and message $m$ - the underlying VIS protocol accepts with probability at least $\alpha$.  
Therefore, the VIQDS protocol also accepts with probability at least $\alpha$, satisfying $\alpha$-completeness.
\end{proofa}

\begin{thm}
If the original quantum VIS protocol is a classical (or quantum) specious-verifier QZKP and satisfies $\beta$-soundness, then the resulting VIQDS protocol is $\epsilon$-information-theoretically unforgeable with classical (or quantum) specious participants, where $\epsilon = \beta$.
\end{thm}

\begin{proofa}
Suppose the original VIS protocol satisfies the assumptions.  
Assume an adversary attempts a forging attack:  
the verifier requests a signature on message $m$,  
but the adversary modifies the pre-signing phase so that the signer is asked to sign a different message $m' \neq m$.

Since all colluding participants are classical (or quantum) specious, they have learned nothing about the signer’s private key — even though they interacted with the signer during the protocol.  
That is,
any colluding participants have to keep their public key that contains the information for private key in their own memory.
This is guaranteed by the specious zero-knowledge property of the VIS protocol.

Now, consider the signer’s behavior:  
she generates a signature based on her private key $sk$ and the message $m'$.  
From the verifier’s perspective, this signature appears to be generated for $m'$, not $m$.  
But since the adversary has no knowledge of the private key, and the VIS protocol satisfies $\beta$-soundness,  
the probability that the verifier accepts this forged signature — even after the adversary’s manipulation — is at most $\beta$.

Thus, the VIQDS protocol is $\beta$-unforgeable against classical (or quantum) specious participants.
\end{proofa}

\subsection{Concatenation of VIS protocols}
To show Theorem \ref{TB1}, we prove several lemmas as follows.
\begin{lem}\label{LN1}
When the original VIS protocols
${\cal V}^1$ and ${\cal V}^2$ 
have 
$\alpha_1$-completeness and 
$\alpha_2$-completeness, respectively,
then the concatenated VIS protocol 
${\cal V}^1\otimes {\cal V}^2$ 
has
$\alpha_1\alpha_2$-completeness,
\end{lem}

\begin{proofa}
Remember that the states on 
${\cal H}_r^1$, ${\cal H}_r^2$ for $r$
are independently prepared.
When both players behave honestly
in the concatenated VIS protocol 
${\cal V}^1\otimes {\cal V}^2$, 
the verifier accepts with the probability 
$\alpha_1\alpha_2$
because the acceptance probability $X_1^1=1$ ($X_1^2=1$) is $\alpha_1$ ($\alpha_2$)
and $X_1^1$ and $X_1^2$ are independent.
Thus, the concatenated VIS protocol 
${\cal V}^1\otimes {\cal V}^2$
has 
$\alpha_1\alpha_2$-completeness.
\end{proofa}

\begin{lem}\label{LN2}
When the original VIS protocols
${\cal V}^1$ and ${\cal V}^2$ 
have 
$\beta_1$-soundness of type 1 and 
$\beta_2$-soundness of type 1, respectively,
then the concatenated VIS protocol 
${\cal V}^1\otimes {\cal V}^2$ 
has
$\beta_1\beta_2$-soundness of type 1.
\end{lem}

\begin{proofa}
When the verifier behaves honestly with $r$
and the prover behaves honestly based on incorrect $r'\neq r$.
The verifier's acceptance probability is upper bounded by
$\beta_1\beta_2$
because the acceptance probability $X_1^1=1$ ($X_1^2=1$) is upper bounded by $\beta_1$ ($\beta_2$)
and $X_1^1$ and $X_1^2$ are independent.
Thus, the concatenated VIS protocol 
${\cal V}^1\otimes {\cal V}^2$
has $\beta_1\beta_2$-soundness of type 1.
\end{proofa}

\begin{lem}\label{LN3}
When the original VIS protocols
${\cal V}^1$ and ${\cal V}^2$ 
have the zero knowledge property $t$ with
$t \in \{\mathsf{HV}, \mathsf{QSV}, \mathsf{CSV}, \mathsf{DV}\}$,
then the concatenated VIS protocol 
${\cal V}^1\otimes {\cal V}^2$ 
has
the zero knowledge property $t$.
\end{lem}

\begin{proofa}
When both protocols 
${\cal V}^1$ and ${\cal V}^2$ 
satisfy the zero-knowledge property $\mathsf{HV}$,
since the states on 
${\cal H}_r^1$, ${\cal H}_r^2$ for $r$
are independently prepared,
the concatenated VIS protocol 
${\cal V}^1\otimes {\cal V}^2$ 
also satisfies the zero-knowledge property $\mathsf{HV}$.

Next, we assume that both protocols 
${\cal V}^1$ and ${\cal V}^2$ 
satisfy the zero-knowledge property $\mathsf{QSV}$.
We consider the final state of the protocol
in the protocol ${\cal V}^1\otimes {\cal V}^2$ with a statement $r$,
and denote the verifier's final system by $X_1$.
We focus on the mutual information
$I(X_1; W_r^1,W_r^2)
=I(X_1; W_r^1)
+I(X_1; W_r^2|W_r^1)$.
We also assume that the prover is honest,
and the verifier is a quantum specious verifier.
Hence, the prover's behavior can be divided into two parts, 
the prover P1 on ${\cal V}^1$ and the prover P2 on ${\cal V}^2$
because their behaviors are independent.
That is, $P1$ is independent of $W_r^2$, and
$P2$ is independent of $W_r^1$.

Now, we focus on the prover P1 and the current verifier 
on the protocol ${\cal V}^1$.
The current verifier is also a quantum specious verifier of 
the protocol ${\cal V}^1$
because the prover P1 detects no difference from 
the honest verifier.
Since ${\cal V}^1$ satisfies the zero-knowledge property $\mathsf{QSV}$,
we have $I(X_1; W_r^1)=0$.

Next, we consider the party V2 composed of the verifier and P1,
which can be considered as a verifier of ${\cal V}^2$.
The verifier V2 is also a quantum specious verifier of 
the protocol ${\cal V}^2$
because the prover P2 detects no difference from 
the honest verifier.
Since ${\cal V}^2$ satisfies the zero-knowledge property $\mathsf{QSV}$,
\begin{align}
0=I(X_1 W_r^1;W_r^2)
=I(W_r^1; W_r^2)
+I(X_1; W_r^2|W_r^1)
=I(X_1; W_r^2|W_r^1).
\end{align}
Therefore, we conclude that
$I(X_1; W_r^1,W_r^2) =0$.
That is,
the concatenated VIS protocol 
${\cal V}^1\otimes {\cal V}^2$ 
satisfies the zero-knowledge property $\mathsf{QSV}$.

Similarly,
when both protocols 
${\cal V}^1$ and ${\cal V}^2$ 
satisfy the zero-knowledge property $\mathsf{CSV}$ ($\mathsf{DV}$),
we can show that
the concatenated VIS protocol 
${\cal V}^1\otimes {\cal V}^2$ 
satisfies the zero-knowledge property $\mathsf{CSV}$ ($\mathsf{DV}$).
\end{proofa}

In addition, when 
$V_1^1$ or $V_1^2$ is a classical system, we have the following lemma.

\begin{lem}\label{LN4}
When the original VIS protocols
${\cal V}^1$ and ${\cal V}^2$ 
have 
$\beta_1$-soundness of type 2 and 
$\beta_2$-soundness of type 2, respectively,
then the concatenated VIS protocol 
${\cal V}^1\otimes {\cal V}^2$ 
has
$\beta_1\beta_2$-soundness of type 2.
\end{lem}

This lemma will be shown in Supplementary Note 3.
The combination of Lemmas \ref{LN1}, \ref{LN2}, \ref{LN3}, and 
\ref{LN4} implies Theorem \ref{TB1}.

\subsection{Construction and security proof of concrete QZKP protocol}
We now present a concrete quantum VIS protocol based on measurements in the discrete Heisenberg representation
and their concatenation.
This protocol serves as the building block for our VIQDS protocol via the general compiler introduced in Results.  
Its security is derived from the algebraic structure of finite fields — a feature that enables tight, information-theoretic guarantees.

We employ the discrete Heisenberg group over the finite field $\mathbb{F}_p$ \cite[Chapter 8]{H-group2}.  
This choice is not arbitrary: it provides a natural framework for constructing non-commuting observables, which are essential for ensuring that a verifier without knowledge of the eigenbasis cannot reliably extract the eigenvalue from a single quantum system.

For $s,t \in \mathbb{F}_p$, let $\mathcal{H}$ be the Hilbert space spanned by the basis
$\{\ket{x}\}_{x \in \mathbb{F}_p}$. We define two operators $\sfX(s)$ and $\sfZ(t)$ on
$\mathcal{H}$ by
\begin{align}
\sfX(s)\ket{x}=\ket{x+s},
\qquad
\sfZ(t)\ket{x}=\omega^{xt}\ket{x},
\end{align}
where $\omega=e^{2\pi i/p}$. Thus, $\sfX(s)$ shifts the basis states by $s$, whereas
$\sfZ(t)$ applies the phase factor $\omega^{xt}$.
The discrete Heisenberg operator $\sfW(s,t) := X(s)Z(t)$ satisfies the commutation relation:
\begin{align}
\sfW(s,t)\sfW(s',t') = \omega^{s't - t's} \sfW(s',t')\sfW(s,t).
\end{align}

For $t,t',s \in  \mathbb{F}_p$,
we define the initial states
$|\phi(t,t')\rangle$ as the eigenvector of 
$\sfW(s,s t)$ 
with eigenvalue $\omega^{ s t'} $.
The eigenvectors of $\sfW(s,s t)$ with eigenvalue $\omega^{s t'} $,
denoted $|\phi(t,t')\rangle$, serve as the witness in our protocol.  
Given a matrix $\sfX=\sum_{k,l \in \mathbb{F}_p}x_{k,l}|k\rangle \langle l|$, we define the vector $|\sfX \rangle\rangle$ on
${\cal H}^{\otimes 2}$ as
\begin{align}
|\sfX \rangle\rangle :=\sum_{k,l \in \mathbb{F}_p}x_{k,l}
|k\rangle |l\rangle.
\end{align}

A key property in our one-copy setting is that, without knowledge of the eigenbasis, a verifier cannot reliably determine the eigenvalue from a single copy of the public-key state.
The protocol proceeds in three rounds, as illustrated in Figure~\ref{F1},
and is denoted by $\mathcal{V}[p]$.
The prover aims to prove knowledge of a witness $W_r = (W_{r,1}, W_{r,2}, W_{r,3}, W_{r,4})$, where $r \in \{1, \ldots, n\}$.  
The public information consists of $2n$ quantum systems prepared in the states $|\phi(W_{r,1},W_{r,2})\rangle$ and $|\phi(W_{r,3},W_{r,4})\rangle$ for each $r$, which are
stored in the verifier's quantum memory $\mathcal M$.

\begin{itemize}
    \item In Step 1,
    the verifier generates a uniform random number $J \in \mathbb{F}_p$ and applies $\sfZ(J) \otimes \sfZ(J)$ to the quantum state $|\phi(W_{r,1},W_{r,2})\rangle \otimes |\phi(W_{r,3},W_{r,4})\rangle$, whose output system is $X_0$. 
Then, the variable $J$ is kept in the system $V_1$.
    \item In Step 2,
    the prover measures the state in the basis $\{|\phi(W_{r,1},k_1)\rangle \otimes |\phi(W_{r,3},k_2)\rangle\}_{k_1,k_2}$ and obtains outcomes $K_1, K_2$.  
    If $W_{r,2} - K_1 = W_{r,4} - K_2$, the prover sends $J' := W_{r,2} - K_1$ to the verifier; otherwise, the protocol terminates.
    That is, the prover sets the variable $J'$ in the system $Y_0$. 
    \item In Step 3,
    the verifier compares $J'$ with $J$, i.e., 
measures the systems $V_1$ and $Y_0$ with the computation basis, and compares their outcomes.
    If they match, it outputs 1 (accept); otherwise, 0 (reject).
\end{itemize}

The protocol requires only four simple operations:  
randomness generation, application of generalized $Z$ gates, a single-shot measurement, and classical comparison.  
Communication is minimal: two classical values and two quantum systems per round.
We now prove the security properties of this protocol.  

\begin{lem}\label{LL1}
The above protocol $\mathcal{V}[p]$ is $QIP_{1,\frac{1}{p}}$.
\end{lem}

This lemma follows directly from the protocol’s structure: an honest prover always passes, while a dishonest prover has at most a $1/p$ chance of guessing the correct value.  
Its formal proof is given in Supplementary Note 1.

\begin{thm}\label{TH1}
The above protocol $\mathcal{V}[p]$ is 
$QZKP^{\mathsf{CSV}}_{1,\frac{1}{p}}$.
\end{thm}

The combination of Theorems \ref{TB1} and \ref{TH1}
implies that
the protocol $\mathcal{V}[p]^{\otimes l}$ is 
$QZKP^{\mathsf{CSV}}_{1,\frac{1}{p^l}}$.
When $p$ is $2$, this protocol can be implemented by qubit technologies.

\begin{remark}[CSV vs.\ QSV for the concrete instantiation]
Our main instantiation is proved to be zero-knowledge against \emph{classical} specious verifiers (CSV) (Theorem~\ref{TH1}).
Since any quantum specious verifier is also a classical specious verifier in our definition hierarchy, CSV zero-knowledge immediately implies QSV zero-knowledge in the one-copy setting considered here.
\end{remark}

\begin{proofa}
To show Theorem \ref{TH1}, we assume that the prover P follows the protocol with the correct witness $W_r$.  
The verifier V’s operation at step 0 is written as an instrument $\{ C_j\}_{j\in {\cal J}}$, where ${\cal J}$ is the set of measurement outcomes.  
That is, $C_j$ is a completely positive map on ${\cal H}_{r,1} \otimes {\cal H}_{r,2}$, and $\sum_{j \in \mathbb{F}_p}C_j$ is a trace-preserving completely positive map on ${\cal H}_{r,1} \otimes {\cal H}_{r,2}$.  
We denote the Choi matrix of $C_j$ by $D_j$.

In this scenario, the set ${\cal J}$ is not necessarily $\mathbb{F}_p$.  
To discuss this case, we introduce an additional notation.  
Using the projection $ p^{-1}|\sfW(0,j)\rangle\rangle\langle\langle \sfW(0,j)|$, we define the projection $\Pi_j$ on ${\cal H}^{\otimes 4}$ as
\begin{align}
\Pi_j:=
p^{-2}
|\sfW(0,j)\rangle\rangle \langle \langle \sfW(0,j)|
\otimes |\sfW(0,j)\rangle\rangle \langle \langle \sfW(0,j)|.
\end{align}
To show Theorem \ref{TH1}, we present the following lemmas.
\begin{lem}\label{LL2}
The verifier V is classical specious if and only if
\begin{align}
\Tr \Big(p^{-2}\sum_{j \in {\cal J}} D_j \Big) 
\Big(\sum_{j' \in \mathbb{F}_p}\Pi_{j'}\Big)
= 1.\label{VXH}
\end{align}
\end{lem}
This lemma is shown in Supplementary Note 2.
\begin{lem}\label{LL3}
The condition \eqref{VXH} implies the relation
\begin{align}
\mathbb{P}(J=j, J'=j' |
(W_{r,1},W_{r,2},W_{r,3},W_{r,4})=(w_{r,1},w_{r,2},w_{r,3},w_{r,4})
)= \Tr p^{-2} D_j \Pi_{j'}
\end{align}
with any choices of $w_{r,1},w_{r,2},w_{r,3},w_{r,4} \in \mathbb{F}_p$.
\end{lem}

Due to Lemmas \ref{LL2} and \ref{LL3}, when the verifier V is classical specious, the probability does not depend on the choices of $w_{r,1},w_{r,2},w_{r,3},w_{r,4} \in \mathbb{F}_p$.  
Therefore, the verifier V obtains no information about the witness $W_r$.  
That is, the protocol is a classical-specious-verifier quantum zero-knowledge proof.

Thus, combining Lemma \ref{LL1},
we conclude that the above protocol is
$QZKP^{\mathsf{CSV}}_{1,\frac{1}{p}}$,
which proves Theorem \ref{TH1}.
\end{proofa}

\begin{proofof}{Lemma \ref{LL3}}
Since the condition \eqref{VXH} holds, we have the following calculation:
\begin{align}
&\mathbb{P}(J=j, J'=j' |(W_{r,1},W_{r,2},W_{r,3},W_{r,4})=(w_{r,1},w_{r,2},w_{r,3},w_{r,4})
)\notag \\
=&
\Tr D_j
\Big[
(|\bar{\phi}(w_{r,1},w_{r,2})\rangle | \bar{\phi}(w_{r,3},w_{r,4})\rangle
\langle \bar{\phi}(w_{r,1},w_{r,2})| \langle \bar{\phi}(w_{r,3},w_{r,4})| )\notag \\
&\otimes
(|\phi(w_{r,1},w_{r,2}+{j'})\rangle | \phi(w_{r,3},w_{r,4}+{j'})\rangle
\langle \phi(w_{r,1},w_{r,2}+{j'})| \langle \phi(w_{r,3},w_{r,4}+{j'})| )\Big] 
\notag\\
=& \Tr p^{-2} D_j \Pi_{j'}.
\end{align}
\end{proofof}

\subsection{Conversion to VIQDS}

The concrete QZKP protocol described above can be systematically converted into a verifier-initiated quantum digital signature (VIQDS) protocol via our general compiler (see Results).  
The resulting VIQDS inherits the security properties of the underlying QZKP, including robustness against dishonest signers and protection against specious verifiers.  
The mapping between the QZKP and VIQDS is summarized in Table~\ref{fig:qzkp_viqds_mapping}.

\subsection{Experimental feasibility under current technologies}
Our proposed protocol is fundamentally constructed by the concatenation of a Quantum Zero-Knowledge Proof (QZKP) protocol based on $p$-dimensional qudit systems. The base QZKP protocol requires only the unitary operations $\sfZ(t)$, projective measurements for the basis $\{|\phi(t,t')\rangle\}_{t'}$, and quantum memory.
While the qudit-based protocol provides the general framework, it is sufficient to combine several qubit-based QZKP protocols to implement a robust security parameter through concatenation. In the case of qubits, the base protocol simplifies, requiring only the unitary operation $\sigma_z$, and projective measurements in the $\sigma_x$ and $\sigma_y$ bases, in addition to quantum memory. These operations can be realized by the combination of the measurement of $\sigma_x$ and the $Z$-axis $\pi/2$ rotation $R_z(\pi/2)$.

Experimentally, implementing these required operations, measurements, and practical quantum memories is increasingly feasible, although platform-specific trade-offs persist. Photonic systems naturally support both qudit and qubit encodings—utilizing degrees of freedom such as time-bins\cite{zheng2022}, orbital angular momentum\cite{kim2024}, or combined polarization-spatial modes\cite{cheng2023}. Recent integrated-photonics work has demonstrated high-fidelity, high-dimensional state preparation and efficient analyzers, making rich qudit gates (as well as qubit gates) and measurements practical on chip-scale hardware\cite{chi2023}. Both qubit and qudit measurements are realized by adopting multiport beamsplitters built with multicore optical fibers\cite{martinez2023}. Photonics' main memory challenge is that flying photons do not self-store; thus, progress focuses on hybrid approaches (e.g., solid-state emitters\cite{zhou2023} coupled to nanophotonic cavities and atomic gases\cite{jing2024m}) and engineered delay techniques\cite{bozkurt2025} that trade complexity for longer-lived photonic storage.

Superconducting platforms, by contrast, can implement qudits (multilevel transmon\cite{liu2023} and cavity modes\cite{reineri2023}) directly, as well as qubits. These platforms have demonstrated dynamical-decoupling and gate protocols that meaningfully suppress decoherence for multilevel operations\cite{han2025}, while cavity memories (3D\cite{weiss2024} or planar high-Q resonators\cite{krayzman2024}) provide some of the longest on-chip storage times available today. Furthermore, superconducting generalized measurement is implemented by a hybrid method called the Naimark-terminated binary tree\cite{ivashkov2024}. However, this method has so far been realized only for superconducting qubits, highlighting the imperative to develop a generalized, experimentally viable measurement protocol for superconducting qudits.

Although our main focus is the ideal setting, we also consider three standard implementation effects when assessing feasibility:
(i) \emph{finite-coherence memory}, where the verifier's one-copy public key undergoes a storage channel $\Lambda_{\mathrm{mem}}(\tau)$ over dwell time~$\tau$;
(ii) \emph{imperfect operations/measurements}, where each intended CPTP map $Z_j$ is implemented as $\tilde Z_j=\mathcal N_j\circ Z_j$ and the verification POVM effect is perturbed within an operator-norm budget; and
(iii) \emph{detection loss}, where the decisive measurement may output an erasure symbol $\bot$ with probability $\lambda_{\mathrm{loss}}$ under abort-on-erasure or reject-on-erasure conventions.
The resulting renormalization of the completeness/soundness parameters and the corresponding (approximate) zero-knowledge guarantee are stated and proved in the Supplementary Note 4 (Robustness Theorem), together with explicit sufficient feasibility thresholds in Supplementary Note 5.

\subsection{Data Availability}
No new datasets were generated or analysed in this study.

\subsection{Acknowledgements}
The authors thank Professor François Le Gall for valuable discussions and helpful suggestions.
W.W. was supported in part by 
Japan Society for the Promotion of Science
Grant in Aid for Scientific Research (S) JP24H00071 and 25K24674,
JST SPRING, Grant Number JPMJSP2125 and the "THERS Make
New Standards Program for the Next Generation Researchers."
M.H. was supported in part by
the General R\&D Projects of 1+1+1 CUHK-CUHK(SZ)-GDST Joint Collaboration Fund (Grant No. GRDP2025-022), the Guangdong Provincial Quantum Science Strategic Initiative (Grant No. 
GDZX2505003), 
the Shenzhen International Quantum Academy (Grant No. SIQA2025KFKT07),
and the National Natural Science Foundation of China under Grant 62171212.

\subsection{Author contributions}

W.W. and M.H. contributed to the conception of the work, the analysis and interpretation of the work, and the preparation of the manuscript.

\subsection{Competing interests}

The authors declare no competing interests.

\begin{table}[htbp]
    \centering
    \begin{tabular}{|c|c|c|c|}
    \hline
Study &Task & Initiation Model & System \\ 
    \hline
\cite{morimae2022,MY22,CM24}& Message-authentication & Signer-initiated & Quantum \\
    \hline
\cite{Zawadzki2018,Chen2023}
& Identity-authentication & Verifier-initiated & Quantum \\
    \hline
\cite{Fang2020,Kumara23,NIST80063A} & Message-authentication& Verifier-initiated & Classical \\
    \hline
This work & Message-authentication  & Verifier-initiated & Quantum \\
    \hline
    \end{tabular}
    \caption{Comparison of initiation models and system types in prior results and this study.}
    \label{fig1}
\end{table}

\begin{table}[htbp]
    \centering
    \begin{tabular}{|c|c|c|}
    \hline
    \textbf{Quantum VIS protocol (QZKP)} & \textbf{Role} & \textbf{Verifier-initiated Quantum Digital Signature (VIQDS)} \\
    \hline
    Prover (P) & $\rightarrow$ & Signer (Alice) \\
    \hline
    Verifier (V) & $\rightarrow$ & Verifier (Bob) \\
    \hline
    Statement ($r$) & $\rightarrow$ & Message ($m$) \\
    \hline
    Witness ($W_r$) & $\rightarrow$ & Private Key ($sk$) \\
    \hline
    Public Information ($\rho_r$) & $\rightarrow$ & Public Key ($pk$) \\
    \hline
    Round 0 (Preparation) & $\rightarrow$ & Pre-signing Phase \\
    \hline
    Round 1 (Verification) & $\rightarrow$ & Verification Phase \\
    \hline
    Response ($Y_0$) & $\rightarrow$ & Signature \\
    \hline
    \end{tabular}
    \caption{Mapping from a VIS protocol to a VIQDS protocol. This table illustrates how the roles and information in the QZKP are translated into the corresponding elements of the VIQDS protocol. This table shows a more precise correspondence than Figure \ref{F3}.}
    \label{fig:qzkp_viqds_mapping}
\end{table}

\newpage

\subsection{Figure legends}

\refstepcounter{figure}\label{F3}
\noindent Figure \thefigure\ | Relations between VIQDS and QZKP. The double-headed arrows indicate the correspondence between the two frameworks. Here, $sk$ denotes the private key, $pk$ the public key, and $sgn$ the signature.

\medskip

\refstepcounter{figure}\label{FX2}
\noindent Figure \thefigure\ | Key generation and pre-signing. 
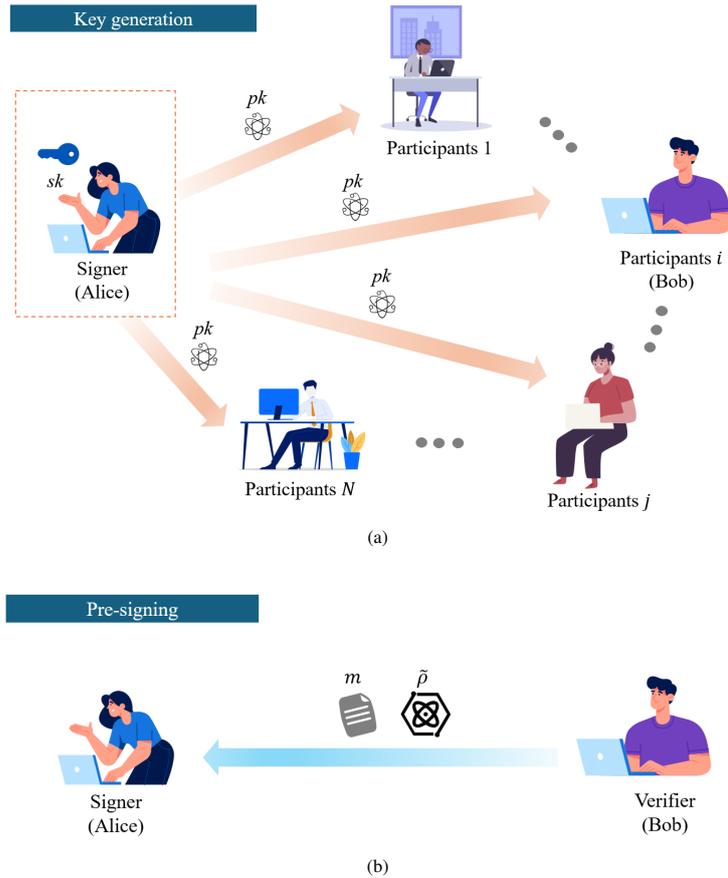
(a) Key generation. Alice generates a private key $sk$ and $N$ copies of the quantum public key $pk$, and sends one copy to each of the remaining participants. 
(b) Pre-signing. Bob requests Alice to sign a message $m$ by sending $m$ together with a quantum challenge $\tilde{\rho}$.

\medskip

\refstepcounter{figure}\label{FX3}
\noindent Figure \thefigure\ | Signing and verification. 
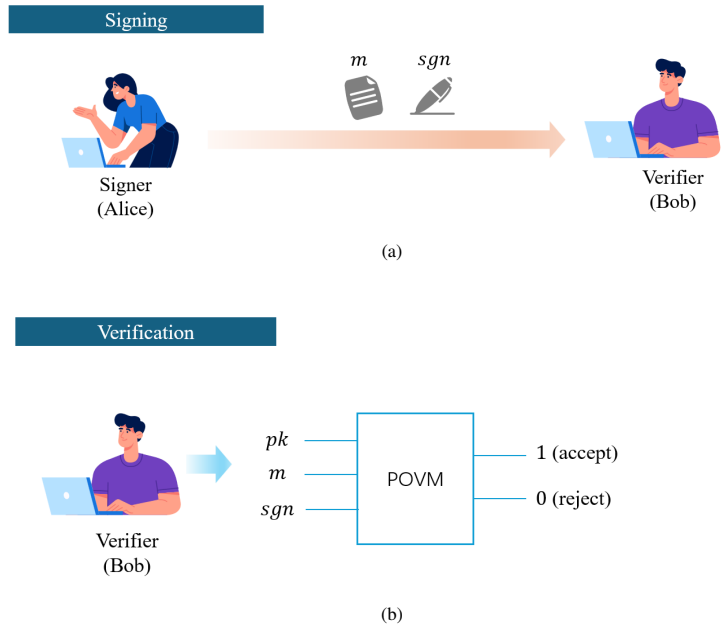
(a) Signing. Alice generates the signature $sgn$ using $sk$ and $\tilde{\rho}$, and then sends the message--signature pair $(m,sgn)$ to Bob. 
(b) Verification. Bob performs a binary POVM using $pk$, $m$, and $sgn$ as inputs. He accepts the signature if the outcome is $1$ and rejects it otherwise.

\medskip

\refstepcounter{figure}\label{F1}
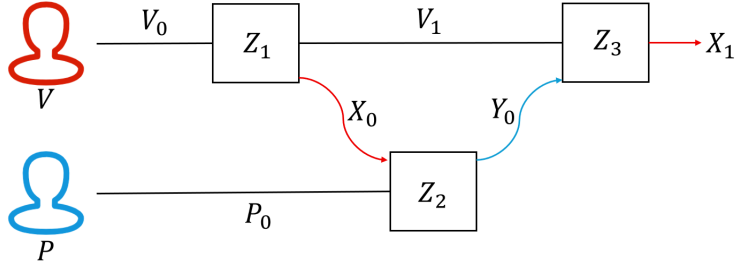
\noindent Figure \thefigure\ | A 3-round quantum interactive proof game. 
$Z_1$, $Z_2$, and $Z_3$ denote the quantum operations performed in the respective steps. 
$P_0$ denotes the prover's internal registers, $V_0$ and $V_1$ denote the verifier's internal registers, $Y_0$ denotes the message sent in round 2, and $X_0$ and $X_1$ denote the messages sent in rounds 1 and 3, respectively.

\medskip

\refstepcounter{figure}\label{F2}
\noindent Figure \thefigure\ | Relationships among $HVQZKP$, $QSVQZKP$, $CSVQZKP$, $DVQZKP$, $QIP$, and the quantum interactive proof game.

\appendix

\section{Proof of Lemma 5}
In this section we prove Lemma~5 in the main paper.
For this aim, we prepare 
the following key lemma for Discrete Heisenberg representation,
which records a simple covariance property of the discrete Heisenberg operators with respect to the eigenbasis used in our construction. 
Informally, acting on an eigenstate of the commuting family $\{\sfW(s,sw_{r,1})\}_s$ with a suitable displacement $\sfW(s,sw_{r,1}-j')$ preserves the label $w_{r,1}$ and shifts the conjugate label $w_{r,2}$ by $j'$ (up to a global phase).  We will use this fact repeatedly to track how the verifier’s random Heisenberg operation transforms the witness states in the completeness and soundness analyses.

\begin{lem}\label{app:PW1}
	Assume that $|{\phi}(w_{r,1},w_{r,2})\rangle$ is the eigenvector of $\sfW(s',s'w_{r,1})$,
	where $w_{r,1},w_{r,2}\in \mathbb{F}_p$.
	For any $s,j' \in \mathbb{F}_p$, we have
	\begin{equation}
 \sfW(s, s w_{r,1}-j')|{\phi}(w_{r,1},w_{r,2})\rangle
 =c_{w_{r,2},s}|\phi(w_{r,1},w_{r,2}+{j'})\rangle,
	\end{equation}
	where $c_{w_{r,2},s}$ is a constant and 
	$|c_{w_{r,2},s}|=1$.
\end{lem}

\begin{proof}
	Since
\begin{align}
&\sfW(s',s'w_{r,1}) \sfW(0,-j') \sfW(s,s w_{r,1}) 
|{\phi}(w_{r,1},w_{r,2})\rangle
=\omega^{tr s' j'}
\sfW(0,-j') \sfW(s,s w_{r,1}) \sfW(s',s'w_{r,1}) 
|{\phi}(w_{r,1},w_{r,2})\rangle \notag\\
=&\omega^{tr s' (j'+w_{r,2})}
\sfW(0,-j') \sfW(s,s w_{r,1}) 
|{\phi}(w_{r,1},w_{r,2})\rangle,
\end{align}
the operator $\sfW(0,-j') \sfW(s, s w_{r,1}) $
maps $|{\phi}(w_{r,1},w_{r,2})\rangle$
to a constant times of $|\phi(w_{r,1},w_{r,2}+{j'})\rangle$
for any $s \in \mathbb{F}_p$. We denote the constant by $c_{w_{r,2},s}$.
Then, the above lemma is obtained.
\end{proof}

Now, we move to the proof of Lemma~5 in the main paper, namely that the base VIS protocol $\mathcal{V}[p]$ achieves perfect completeness and $1/p$ soundness, i.e., it is a $QIP_{1,1/p}$.

For completeness, we analyze the honest execution and show that the verifier always accepts when the prover holds the correct witness.
For soundness, we treat separately (Type~1) an honest-but-wrong witness and (Type~2) a witness-independent (dishonest) prover strategy, and we upper-bound the acceptance probability in both cases by $1/p$.

\begin{lem}\label{app:LL1}
The above protocol $\mathcal{V}[p]$ is $QIP_{1,\frac{1}{p}}$.
\end{lem}

\begin{proof}
{\bf Step 1: Preparation}\quad 
Assume that the verifier V is honest.
The state $\sfX_0$ is
\begin{align}
\sfZ(J)|\phi(W_{r,1},W_{r,2})\rangle \sfZ(J)| \phi(W_{r,3},W_{r,4})\rangle
=
e^{i \theta} |\phi(W_{r,1},W_{r,2}-J)\rangle | \phi(W_{r,3},W_{r,4}-J)\rangle,
\end{align}
where we apply Lemma \ref{app:PW1} and $e^{i \theta}$ is a phase factor.

{\bf Step 2: Completeness}\quad 
When the prover P follows the protocol with the correct witness $W_r$, we have
\begin{align}
W_{r,2}-K_{1}&= W_{r,2}-(W_{r,2}-J)=J \\
W_{r,4}-K_{2}&= W_{r,2}-(W_{r,4}-J)=J .
\end{align}
Hence, the prover P obtains 
$W_{r,2}-K_{1}=W_{r,4}-K_{2}$, and sends
$J'=W_{r,2}-K_{1}=J$ to the verifier V.
Then, the verifier V outputs 1. 
Thus, the protocol satisfies $1$-completeness.

{\bf Step 3: Soundness of type 1}\quad 
Assume that the prover P follows the protocol with the incorrect witness $W_r'=(W_{r,1}',W_{r,2}',W_{r,3}',W_{r,4}')\neq W_r=(W_{r,1},W_{r,2},W_{r,3},W_{r,4})$.
Assume that $W_{r,1}'\neq W_{r,1}$.
Due to the mutually unbiased property of the basis,
the conditional distribution
$P_{K_{1}|W_{r,2},J} $ is the uniform distribution 
regardless of the condition.
Hence, $P_{K_{1}-W_{r,2}'|W_{r,2},J} $ is also the uniform distribution
regardless of the condition.
Thus, $P_{K_{1}-W_{r,2}'|J} $ is also the uniform distribution
regardless of the condition.
Hence, the probability of holding the relation 
$J=J'$ is probability $p^{-1}$.
Thus, 
\begin{align}
\mathbb{P} (J=J', W_{r,2}-K_{1}=W_{r,4}-K_{2})
\le \mathbb{P} (J=J') \le p^{-1}.
\end{align}
When $W_{r,3}'\neq W_{r,3}$, we can derive the same conclusion 
as the above.
We assume that $W_{r,1}'\neq W_{r,1}$ and $W_{r,3}'\neq W_{r,3}$.
When $W_{r,2}'\neq W_{r,2}$,
we have
\begin{align}
W_{r,2}-K_{1}=W_{r,2}-(W'_{r,2}-J)\neq J,
\end{align}
which implies 
\begin{align}
\mathbb{P} (J=J', W_{r,2}-K_{1}=W_{r,4}-K_{2})
\le \mathbb{P} (J=J') =0.
\end{align}
When $W_{r,4}'\neq W_{r,4}$,
we have
\begin{align}
W_{r,4}-K_{2}=W_{r,4}-(W'_{r,4}-J) \neq J,
\end{align}
which implies 
\begin{align}
\mathbb{P} (J=J', W_{r,2}-K_{1}=W_{r,4}-K_{2})
\le \mathbb{P} (J=W_{r,4}-K_{2}) =0
\end{align}
because $J'= W_{r,2}-K_{1}$.
Thus, the protocol satisfies $p^{-1}$-soundness of type 1.

{\bf Step 4: Soundness of type 2}\quad 
Assume that the prover P follows
a deviate protocol that depends on $r$ and is independent of $W_r$.
P's outcome can be considered as $\mathbb{F}_p^* :=\mathbb{F}_p \cup\{ *\}$, where $*$ expresses the event of the condition
$ W_{r,2}-K_{1}\neq W_{r,4}-K_{2}$. 
Hence, P's protocol can be written as a POVM 
$\{ M_{j'}\}_{j' \in \mathbb{F}_p^*}$
on ${\cal H}_{r,1} \otimes {\cal H}_{r,2}$. 
We define $\tilde{W}_{r,2}:=W_{r,2}-J$ and $\tilde{W}_{r,4}:=W_{r,4}-J$.
We have
\begin{align}
& \mathbb{P} (J=J', K_{1}-W_{r,2}= K_{2}-W_{r,4}) \notag\\
=&
\mathbb{E}[\sum_{j' \in \mathbb{F}_p: j'=J } 
\langle \phi(W_{r,1},W_{r,2}-J) | \langle \phi(W_{r,3},W_{r,4}-J)| M_{j'}
|\phi(W_{r,1},W_{r,2}-J)\rangle | \phi(W_{r,3},W_{r,4}-J)\rangle] \notag\\
= &
\mathbb{E}[
\langle \phi(W_{r,1},W_{r,2}-J) | \langle \phi(W_{r,3},W_{r,4}-J)| M_{J}
|\phi(W_{r,1},W_{r,2}-J)\rangle | \phi(W_{r,3},W_{r,4}-J)\rangle] \notag\\
= &
\mathbb{E}[
\langle \phi(W_{r,1},\tilde{W}_{r,2})| \langle \phi(W_{r,3},\tilde{W}_{r,4})| M_{J}
|\phi(W_{r,1},\tilde{W}_{r,2})\rangle | \phi(W_{r,3},\tilde{W}_{r,4})\rangle] \notag\\
= &
\Tr 
\mathbb{E}[M_{J}
|\phi(W_{r,1},\tilde{W}_{r,2})
\langle \phi(W_{r,1},\tilde{W}_{r,2})| \langle \phi(W_{r,3},\tilde{W}_{r,4})| M_{J}
|\phi(W_{r,1},\tilde{W}_{r,2})\rangle | \phi(W_{r,3},\tilde{W}_{r,4})\rangle] \notag\\
= &
\Tr 
\mathbb{E}[M_{J}
|\phi(W_{r,1},\tilde{W}_{r,2})\rangle 
| \phi(W_{r,3},\tilde{W}_{r,4})\rangle\langle \phi(W_{r,1},\tilde{W}_{r,2})| \langle \phi(W_{r,3},\tilde{W}_{r,4})| 
]\notag \\
= &
\Tr \mathbb{E}_J[M_{J}]
\mathbb{E}_{W_{r,1},\tilde{W}_{r,2},W_{r,3},\tilde{W}_{r,4}}
[|\phi(W_{r,1},\tilde{W}_{r,2})\rangle | \phi(W_{r,3},\tilde{W}_{r,4})\rangle\langle \phi(\sfW_{r,1},\tilde{\sfW}_{r,2})| 
\langle \phi(W_{r,3},\tilde{W}_{r,4})| ] \notag\\
= &
\Tr p^{-1}I\cdot p^{-2}I
=q^2 \cdot p^{-1}\cdot p^{-2}=p^{-1}.
\end{align}
Thus, the protocol satisfies $p^{-1}$-soundness of type 2.
Therefore, the above protocol is a $QIP_{1,\frac{1}{p}}$.
\end{proof}

\section{Proof of Lemma 6}
In this section we prove Lemma~6 in the main paper, i.e., the CSV (classical-specious-verifier) condition for the concrete instantiation.
We express the verifier's Step~1 operation as an instrument $\{C_j\}_{j\in\mathcal J}$ with Choi operators $\{D_j\}$, and characterize classical speciousness by requiring that the verifier's induced statistics on the prover's classical memory agree with those of the honest-verifier interaction for every witness value.
This requirement is equivalent to the following trace condition, stated in terms of the projectors $\{\Pi_{j'}\}_{j'\in\mathbb F_p}$ introduced in the main text.
\begin{lem}\label{app:LL2}
The verifier V is classical specious if and only if
\begin{align}
\Tr \Big(p^{-2}\sum_{j \in {\cal J}} D_j \Big) 
\Big(\sum_{j' \in \mathbb{F}_p}\Pi_{j'}\Big)
= 1.\label{app:VXH}
\end{align}
\end{lem}

\begin{proof}
{\bf Step 1: Preparation}\quad 
Since $\Tr p^{-2}\sum_{j \in {\cal J}} D_j=1$, the matrix $\sum_{j \in {\cal J}} D_j $ can be considered as a density matrix. Also, $\sum_{j' \in \mathbb{F}_p} \Pi_{j'}$ is a projection with rank $q$. Hence, the condition is equivalent to 
\begin{align}
\Big(\sum_{j' \in \mathbb{F}_p} \Pi_{j'}\Big)
\Big(\sum_{j \in {\cal J}} D_j\Big)
\Big(\sum_{j' \in \mathbb{F}_p} \Pi_{j'}\Big)
=\sum_{j \in \mathbb{F}_p} D_j.
\end{align}

The verifier V is classical specious if and only if
\begin{align}
1=&\mathbb{P}(K_{1}-W_{r,2}= K_{2}-W_{r,4}
 |(W_{r,1},W_{r,2},W_{r,3},W_{r,4})=(w_{r,1},w_{r,2},w_{r,3},w_{r,4})
)\notag \\
=&
\Tr (\sum_{j \in {\cal J}} D_j)
\sum_{j' \in \mathbb{F}_p}
\Big[
\big(|\bar{\phi}(w_{r,1},w_{r,2})\rangle | \bar{\phi}(w_{r,3},w_{r,4})\rangle
\langle \bar{\phi}(w_{r,1},w_{r,2})| \langle \bar{\phi}(w_{r,3},w_{r,4})| \big)\notag \\
&\otimes
\big(|\phi(w_{r,1},w_{r,2}-{j'})\rangle | \phi(w_{r,3},w_{r,4}-{j'})\rangle
\langle \phi(w_{r,1},w_{r,2}-{j'})| \langle \phi(w_{r,3},w_{r,4}-{j'})| \big)\Big] 
\label{app:VXH1}
\end{align}
for any $w_{r,1},w_{r,2},w_{r,3},w_{r,4} \in \mathbb{F}_p$.

In the above relations, we set $j'$ as the observed outcome of $K_{1}-W_{r,2}= K_{2}-W_{r,4}$.

{\bf Step 2: \eqref{app:VXH}$\Rightarrow$\eqref{app:VXH1}}\quad 
Since
\begin{align}
| \langle \langle \sfW(0,j')|
|\phi(w_{r,1},w_{r,2}-{j'})\rangle|\bar{\phi}(w_{r,1},w_{r,2})\rangle |^2
=1,
\end{align}
the projection $p^{-1}|\sfW(0,j')\rangle\rangle \langle \langle \sfW(0,j')|$ satisfies
\begin{align}
&
q^{-1}|\sfW(0,j')\rangle\rangle \langle \langle \sfW(0,j')|
\Big[
|\phi(w_{r,1},w_{r,2}-{j'})\rangle|\bar{\phi}(w_{r,1},w_{r,2})\rangle 
\langle \phi(w_{r,1},w_{r,2}-{j'})|\bar{\phi}(w_{r,3},w_{r,4})|\Big]\notag \\
&\cdot
q^{-1}|\sfW(0,j')\rangle\rangle \langle \langle \sfW(0,j')|
\notag \\
=&
q^{-1} \Big(q^{-1}|\sfW(0,j')\rangle\rangle \langle \langle \sfW(0,j')|\Big)
 \label{app:TT2}.
\end{align}
Thus,
the projection $\Pi_{j'}$ satisfies
\begin{align}
&
\Pi_{j'}
\Big[
|\phi(w_{r,1},w_{r,2}-{j'})\rangle|\bar{\phi}(w_{r,1},w_{r,2})\rangle 
\langle \phi(w_{r,1},w_{r,2}-{j'})| \langle\bar{\phi}(w_{r,3},w_{r,4})|
\notag \\
&\otimes
| \phi(w_{r,3},w_{r,4}-{j'})\rangle| \bar{\phi}(w_{r,3},w_{r,4})\rangle
 \langle \phi(w_{r,3},w_{r,4}-{j'})| )
\langle \bar{\phi}(w_{r,1},w_{r,2})| 
\Big]
\Pi_{j'}
\notag \\
=&
p^{-2}\Pi_{j'}.
\label{app:VXH2}
\end{align}

When the condition \eqref{app:VXH} holds, we have
\begin{align}
&\Tr \Big(\sum_{j \in {\cal J}} D_j\Big)
\sum_{j' \in \mathbb{F}_p}
\Big[
|\bar{\phi}(w_{r,1},w_{r,2})\rangle | \bar{\phi}(w_{r,3},w_{r,4})\rangle
\langle \bar{\phi}(w_{r,1},w_{r,2})| \langle \bar{\phi}(w_{r,3},w_{r,4})| \notag \\
&\otimes
|\phi(w_{r,1},w_{r,2}-{j'})\rangle | \phi(w_{r,3},w_{r,4}-{j'})\rangle
\langle \phi(w_{r,1},w_{r,2}-{j'})| \langle \phi(w_{r,3},w_{r,4}-{j'})| \Big] 
\notag\\
=&
\Tr
\Big(\sum_{j'' \in {\cal J}} \Pi_{j''}\Big)
\Big(\sum_{j \in {\cal J}} D_j\Big)
\Big(\sum_{j'' \in {\cal J}} \Pi_{j''}\Big)
\notag \\
&\cdot\sum_{j' \in \mathbb{F}_p}
\Big[
|\phi(w_{r,1},w_{r,2}-{j'})\rangle|\bar{\phi}(w_{r,1},w_{r,2})\rangle 
\langle \phi(w_{r,1},w_{r,2}-{j'})| \langle \bar{\phi}(w_{r,3},w_{r,4})|
\notag \\
&\otimes
| \phi(w_{r,3},w_{r,4}-{j'})\rangle| \bar{\phi}(w_{r,3},w_{r,4})\rangle
 \langle \phi(w_{r,3},w_{r,4}-{j'})| 
\langle \bar{\phi}(w_{r,1},w_{r,2})| 
\Big] \notag \\
=&
\Tr
(\sum_{j \in {\cal J}} D_j)
\sum_{j' \in \mathbb{F}_p}
\Pi_{j'}
\Big[
|\phi(w_{r,1},w_{r,2}-{j'})\rangle|\bar{\phi}(w_{r,1},w_{r,2})\rangle 
\langle \phi(w_{r,1},w_{r,2}-{j'})|\langle \bar{\phi}(w_{r,3},w_{r,4})|
\notag \\
&\otimes
| \phi(w_{r,3},w_{r,4}-{j'})\rangle| \bar{\phi}(w_{r,3},w_{r,4})\rangle
 \langle \phi(w_{r,3},w_{r,4}-{j'})| \langle \bar{\phi}(w_{r,1},w_{r,2})| 
\Big]
\Pi_{j'}
\notag \\
\stackrel{(a)}{=}&
\Tr
\Big(\sum_{j \in {\cal J}} D_j\Big)
\sum_{j' \in \mathbb{F}_p}
p^{-2} \Pi_{j'}= 1,
\end{align}
where $(a)$ follows from the above relation \eqref{app:VXH2}.

Thus, the condition \eqref{app:VXH} implies the desired property \eqref{app:VXH1}.

{\bf Step 3: \eqref{app:VXH1}$\Rightarrow$\eqref{app:VXH}}\quad 
Next, we show the opposite direction. The condition \eqref{app:VXH1} implies 
\begin{align}
&p^{-2}
\sum_{w_{r,2},w_{r,4}\in \mathbb{F}_p}
\sum_{j,j' \in \mathbb{F}_p}
\Tr D_j \Big[
(|\phi(w_{r,1},w_{r,2})\rangle | \phi(w_{r,3},w_{r,4})\rangle
\langle \phi(w_{r,1},w_{r,2})| \langle \phi(w_{r,3},w_{r,4})| )^T\notag \\
&\otimes
(|\phi(w_{r,1},w_{r,2}-{j'})\rangle | \phi(w_{r,3},w_{r,4}-{j'})\rangle
\langle \phi(w_{r,1},w_{r,2}-{j'})| \langle \phi(w_{r,3},w_{r,4}-{j'})| )\Big]\notag \\
=& 1
\label{app:VXH5}
\end{align}
for any $w_{r,1},w_{r,3}\in \mathbb{F}_p$.

According to Lemma \ref{app:PW1}, the operator $\sfW(s, s w_{r,1}-j') $ is written as
\begin{align}
\sfW(s, s w_{r,1}-j') 
=
\sum_{w_{r,2} \in \mathbb{F}_p} c_{w_{r,2},s}
|\phi(w_{r,1},w_{r,2}+{j'})\rangle\langle {\phi}(w_{r,1},w_{r,2})|,
\end{align}
where $c_{w_{r,2},s}$ is a complex number with $|c_{w_{r,2},s}|=1$.

Thus,
\begin{align}
| \sfW(s, s w_{r,1}-j') \rangle\rangle
=
\sum_{w_{r,2} \in \mathbb{F}_p} c_{w_{r,2},s}
|\phi(w_{r,1},w_{r,2}+{j'})\rangle| \bar{\phi}(w_{r,1},w_{r,2})\rangle.
\label{app:VXH3}
\end{align}

Since $\{|\sfW(s, s w_{r,1}-j') \rangle\rangle\}_{s}$ are orthogonal to each other, the form \eqref{app:VXH3} guarantees 
\begin{align}
&p^{-2}\sum_{s }| \sfW(s, s w_{r,1}-j') \rangle\rangle 
\langle \langle \sfW(s, s w_{r,1}-j') | \notag \\
=&p^{-1}
\sum_{w_{r,2} \in \mathbb{F}_p} 
|\phi(w_{r,1},w_{r,2}+{j'})\rangle |\bar{\phi}(w_{r,1},w_{r,2})\rangle
\langle \phi(w_{r,1},w_{r,2}+{j'})|\langle \bar{\phi}(w_{r,1},w_{r,2})|
\notag \\
=&p^{-1} \sum_{w_{r,2}\in \mathbb{F}_p}
(|\phi(w_{r,1},w_{r,2})\rangle \langle \phi(w_{r,1},w_{r,2})| )^T
\otimes|\phi(w_{r,1},w_{r,2}+{j'})\rangle\langle \phi(w_{r,1},w_{r,2}+{j'})|.
\label{app:VXH4}
\end{align}

We define the projection
\begin{align}
&\Pi(w_{r,1},w_{r,3} )\notag\\
:=& p^{-2}\sum_{a,b,a'b' : b-a w_{r,1}=b'-a' w_{r,3}}
|\sfW(a,b)\rangle\rangle \langle \langle \sfW(a,b)|
\otimes |\sfW(a',b')\rangle\rangle \langle \langle \sfW(a',b')|
\label{app:VB3}\\
=
& \sum_{j' \in \mathbb{F}_p}
\Big(p^{-1}\sum_{s }|\sfW(s, s w_{r,1}-j') \rangle\rangle 
\langle \langle \sfW(s, s w_{r,1}-j') | \Big) \notag\\
& \otimes \Big(p^{-1}\sum_{s' }| \sfW(s', s' w_{r,3}-j') \rangle\rangle 
\langle \langle \sfW(s', s' w_{r,3}-j') | \Big) 
\notag\\
\stackrel{(a)}{=}
&
\sum_{w_{r,2},w_{r,4}\in \mathbb{F}_p}\sum_{j' \in \mathbb{F}_p}\Big[
(|\phi(w_{r,1},w_{r,2})\rangle | \phi(w_{r,3},w_{r,4})\rangle
\langle \phi(w_{r,1},w_{r,2})| \langle \phi(w_{r,3},w_{r,4})| )^T\notag \\
&\otimes
(|\phi(w_{r,1},w_{r,2}+{j'})\rangle | \phi(w_{r,3},w_{r,4}+{j'})\rangle
\langle \phi(w_{r,1},w_{r,2}+{j'})| \langle \phi(w_{r,3},w_{r,4}+{j'})| )\Big] 
\label{app:VBY},
\end{align}
where $(a)$ follows from the above relation \eqref{app:VXH4}.

Thus, the relation \eqref{app:VXH5} implies 
\begin{align}
& 
\Tr \Big(p^{-2}\sum_{j \in \mathbb{F}_p} D_j \Big) \Pi(w_{r,1},w_{r,3} )\notag\\
\stackrel{(a)}{=} &p^{-2}
\sum_{w_{r,2},w_{r,4}\in \mathbb{F}_p}
\sum_{j,j' \in \mathbb{F}_p}
\Tr D_j 
\Big[
(|\phi(w_{r,1},w_{r,2})\rangle | \phi(w_{r,3},w_{r,4})\rangle
\langle \phi(w_{r,1},w_{r,2})| \langle \phi(w_{r,3},w_{r,4})| )^T\notag \\
&\otimes
(|\phi(w_{r,1},w_{r,2}+{j'})\rangle | \phi(w_{r,3},w_{r,4}+{j'})\rangle
\langle \phi(w_{r,1},w_{r,2}+{j'})| \langle \phi(w_{r,3},w_{r,4}+{j'})| )\Big] \notag\\
\stackrel{(b)}{=}& 1 \label{app:VXH6}
\end{align}
for any $w_{r,1},w_{r,3}\in \mathbb{F}_p$,
where $(a)$ follows from \eqref{app:VBY},
and $(b)$ follows from the above relation \eqref{app:VXH5}.

Since the projections $\{\Pi(w_{r,1},w_{r,3} )\}_{w_{r,1},w_{r,3}\in \mathbb{F}_p}$ are orthogonal to each other, 
the conditions $b=b'$ and $a=a'=0$ hold
if and only of the relation $b-a w
_{r,1}=b'-a' w_{r,3}$ holds
for any $w_{r,1},w_{r,3}\in  \mathbb{F}_p$.
Thus, the definition 
\eqref{app:VB3} implies 
\begin{align}
\Pi(w_{r,1},w_{r,3} ) 
= p^{-2}\sum_{j' \in \mathbb{F}_p}
|\sfW(0,j')\rangle\rangle \langle \langle \sfW(0,j')|
\otimes |\sfW(0,j')\rangle\rangle \langle \langle \sfW(0,j')|.
\label{app:MB1}
\end{align}
Due to this relation \eqref{app:MB1}, 
the condition \eqref{app:VXH6} yields the desired relation \eqref{app:VXH}.
That is, the condition \eqref{app:VXH1} implies the desired property \eqref{app:VXH}.
\end{proof}

\section{Stateless Prover–Verifier game: Proof of Lemma 4}\label{app:Ap2}
\subsection{Main part of the proof of Lemma 4}
In this section we prove the concatenation soundness claim (Lemma~4 in the main paper) for Type~2 soundness.
The key idea is to isolate the soundness analysis from any witness-dependent behavior: a Type~2 dishonest prover,
by definition, has no access to a valid witness and thus can be modeled as a \emph{stateless} channel that acts only on the current incoming message.
We therefore reduce the analysis to a single-round prover--verifier game, and then show that the optimal acceptance probability
is multiplicative under tensor-product composition of two such games.

To make the statement precise, we restate the claim in the present notation:
\begin{lem}\label{app:LN4}
When the original VIS-protocols
${\cal V}^1$ and ${\cal V}^2$ 
have 
$\beta_1$-soundness of type 2 and 
$\beta_2$-soundness of type 2, respectively,
then the concatenated VIS protocol 
${\cal V}^1\otimes {\cal V}^2$ 
has
$\beta_1\beta_2$-soundness of type 2.
\end{lem}

For the above aim,
we consider a single-round interaction between a verifier and a prover
because a prover without witness $\mathcal{W}_r$ can be regarded as a stateless prover.
The verifier prepares a fixed state $\rho_{RA}$ on systems $R$ (reference) and $A$ (channel input). The verifier keeps $R$ and sends system $A$ to the prover. The prover has no prior information, no pre-shared entanglement with the verifier, and no persistent memory; its only action is to implement a quantum channel, 
a completely positive trace-preserving (CPTP) map, 
\[
\mathcal{E}:\mathcal{L}(\mathcal{H}_A)\to\mathcal{L}(\mathcal{H}_B),
\]
which maps the system $A$ to an output system $B$ that is returned to the verifier.

\subsubsection*{Verifier strategy.}
The verifier's strategy consists of the choice of the preparation $\rho_{RA}$ and a binary measurement (POVM) $\{\Pi_{\mathrm{acc}},\Pi_{\mathrm{rej}}\}$ acting on $R\otimes B$. After receiving $B$ the verifier performs this joint measurement and outputs ``Yes'' (accept) or ``No'' (reject),
where $\Pi_{\mathrm{rej}}=I-\Pi_{\mathrm{acc}}$.
Therefore, a strategy of the verifier 
on systems $(\cH_A,\cH_B,\cH_R)$
is written as a pair $(\rho_{RA}, \Pi_{\mathrm{acc}})$.

\subsubsection*{Interaction (protocol).}
\begin{enumerate}
  \item The verifier prepares $\rho_{RA}$ and sends system $A$ to Prover.
  \item The prover applies channel $\mathcal{E}$ to $A$ and returns system $B$ to the verifier.
  \item The verifier measures $R\otimes B$ with $\{\Pi_{\mathrm{acc}},\Pi_{\mathrm{rej}}\}$ and outputs the result.
\end{enumerate}

\subsubsection*{Acceptance probability.}
For a given prover channel $\mathcal{E}$, the acceptance probability is

\begin{align}
p[\rho_{RA}, \Pi_{\mathrm{acc}}](\mathcal{E})
=\Tr \Big[\,\Pi_{\mathrm{acc}}
(\mathcal{I}_R\otimes\mathcal{E})(\rho_{RA} )
\Big],
\end{align}
where $\mathcal{I}_R$ denotes the identity channel on the reference system $R$.
The maximum acceptance probability 
$p[\rho_{RA}, \Pi_{\mathrm{acc}}]$
is defined as
\begin{align}
p_{\max}[\rho_{RA}, \Pi_{\mathrm{acc}}]
:= \max_{\mathcal{E}}
p[\rho_{RA}, \Pi_{\mathrm{acc}}](\mathcal{E})
\end{align}

We introduce the Choi matrix of $\mathcal{E}$ 
on the system $\cH_A \otimes \cH_B$ as
\begin{align}
C[\mathcal{E}]:= \mathcal{E} \otimes \mathcal{I} 
\Big( \Big(\sum_i |ii\rangle\Big) \Big(\sum_j \langle jj|\Big)\Big),
\end{align}
%by $C[\mathcal{E}]$ 
on the system $\cH_A \otimes \cH_B$, 
which satisfies the condition;
\begin{align}
\Tr_B C[\mathcal{E}]=I_A,~ C[\mathcal{E}]\ge 0.
\end{align}
Then, we have
\begin{align}
(\mathcal{I}_R\otimes\mathcal{E})(\rho_{RA} )
=\Tr_A (C[\mathcal{E}] \otimes I_R ) (\rho_{RA}^{T_A}\otimes I_B).
\end{align}
Then, we define the operator 
$G[\rho_{RA}, \Pi_{\mathrm{acc}}]:= 
\Tr_R (\rho_{RA}^{T_A}\otimes I_B)
(\Pi_{\mathrm{acc}}\otimes I_A)$,
where $T_A$ presents the partial transpose on the system $A$.
Then, the probability 
$p[\rho_{RA}, \Pi_{\mathrm{acc}}]
(\mathcal{E})$ is rewritten as
\begin{align}
p[\rho_{RA}, \Pi_{\mathrm{acc}}](\mathcal{E})
=&\Tr_{RB} \Big[ \Pi_{\mathrm{acc}}
[\Tr_A (C[\mathcal{E}]\otimes I_R)
 (\rho_{RA}^{T_A}\otimes I_B)]
\Big]\notag\\
=&\Tr_{RBA} \Big[ (\Pi_{\mathrm{acc}}\otimes I_A)
((C[\mathcal{E}]\otimes I_R)
 (\rho_{RA}^{T_A}\otimes I_B))
\Big]\notag\\
=&\Tr_{AB} \Big[ 
[\Tr_R (\rho_{RA}^{T_A}\otimes I_B)
(\Pi_{\mathrm{acc}}\otimes I_A)]
C[\mathcal{E}]
\Big]
=\Tr_{AB} G[\rho_{RA}, \Pi_{\mathrm{acc}}] C[\mathcal{E}].
\end{align}
Thus, due to the duality in
linear programming with general cone summarized in Subsection \ref{app:Ap2-conic},
the maximum acceptance probability 
$p_{\max}[\rho_{RA}, \Pi_{\mathrm{acc}}]$
is rewritten as
\begin{align}
p_{\max}[\rho_{RA}, \Pi_{\mathrm{acc}}]=
\max_{C\ge 0}
\{\Tr_{AB} G[\rho_{RA}, \Pi_{\mathrm{acc}}] C
| \Tr_B C=I_A \}\label{app:BNJ}
\end{align}

\begin{lem}\label{app:VL0}
We have $G[\rho_{RA}, \Pi_{\mathrm{acc}}]\ge 0$.
\end{lem}

When pairs 
$(\rho_{R1A1}, \Pi_{\mathrm{acc},1})$ and 
$(\rho_{R2A2}, \Pi_{\mathrm{acc},2})$ are 
strategies of the verifier
on systems $(\cH_{A1},\cH_{B1},\cH_{R1})$
and $(\cH_{A2},\cH_{B2},\cH_{R2})$, respectively,
$(\rho_{R1A1}\otimes \rho_{R2A2}, 
\Pi_{\mathrm{acc},1}\otimes \Pi_{\mathrm{acc},2})$ 
forms a strategy 
on the systems
$(\cH_{A1}\otimes\cH_{A2},\cH_{B1}\otimes\cH_{B2},
\cH_{R1}\otimes\cH_{R2})$.

\begin{lem}\label{app:VL1}
The relation
\begin{align}
p_{\max}[\rho_{R1A1}\otimes \rho_{R2A2}, 
\Pi_{\mathrm{acc},1}\otimes \Pi_{\mathrm{acc},2}]
=
p_{\max}[\rho_{R1A1}, \Pi_{\mathrm{acc},1}]
p_{\max}[\rho_{R2A2}, \Pi_{\mathrm{acc},2}]\label{app:safsda1}.
\end{align}
holds.
\end{lem}

Lemma \ref{app:VL1} shows Lemma 4 as follows.
Assume that 
the original VIS-protocols
${\cal V}^1$ and ${\cal V}^2$ 
have 
$\beta_1$-soundness of type 2 and 
$\beta_2$-soundness of type 2, respectively.
We denote the verifier's joint state of ${\cal V}^j$
after the application of $Z_0^j$ by $\rho_{RjAj}$, 
and denote the the verifier's final measurement 
of ${\cal V}^j$
by $\{\Pi_{\mathrm{acc},j}, I-\Pi_{\mathrm{acc},j}\}$,
for $j=1,2$.
Thus, $p_{\max}[\rho_{RjAj}, \Pi_{\mathrm{acc},j}]
= \beta_j$ for $j=1,2$.
Lemma \ref{app:VL1} guarantees that
the acceptance probability is upper bounded by 
$\beta_1\beta_2$
because the prover's behavior is independent of
the witness $(W_r^1,W_r^2)$, which shows the statement of 
Lemma 4.

\subsection{Proof of Lemma \ref{app:VL1}}
The part $\ge$ of \eqref{app:safsda1} is shown as follows.
\begin{align}
&p_{\max}[\rho_{R1A1}\otimes \rho_{R2A2}, 
\Pi_{\mathrm{acc},1}\otimes \Pi_{\mathrm{acc},2}] \notag\\
=&
\max_{C \ge 0}
\{\Tr_{A1A2B1B2} 
(G[\rho_{R1A1}\otimes \rho_{R2A2}, \Pi_{\mathrm{acc}1} \otimes
 \Pi_{\mathrm{acc}2}] )
C
| \Tr_{B1B2} C=I_{A1A2} \} \notag\\
=&
\max_{C \ge 0}
\{\Tr_{A1A2B1B2} 
(G[\rho_{R1A1}, \Pi_{\mathrm{acc}1}] \otimes
G[\rho_{R2A2}, \Pi_{\mathrm{acc}2}] )
C
| \Tr_{B1B2} C=I_{A1A2} \} \notag\\
\ge &
\max_{C_1\ge 0,C_2\ge 0}
\{\Tr_{A1A2B1B2} 
(G[\rho_{R1A1}, \Pi_{\mathrm{acc}1}] \otimes
G[\rho_{R2A2}, \Pi_{\mathrm{acc}2}] )
(C_1\otimes C_2)
| \Tr_{B1} C_1=I_{A1} , \Tr_{B2} C_2=I_{A2}\} \notag\\
=& \max_{C_1\ge 0}
\{\Tr_{A1B1} G[\rho_{R1A1}, \Pi_{\mathrm{acc}1}] C_1
| \Tr_{B1} C_1=I_{A1} \}
\max_{C_2\ge 0}
\{\Tr_{A2B2} G[\rho_{R2A2}, \Pi_{\mathrm{acc},2}] C_2
| \Tr_{B2} C_2=I_{A2} \} \notag\\
=&p_{\max}[\rho_{R1A1}, \Pi_{\mathrm{acc},1}]
p_{\max}[\rho_{R2A2}, \Pi_{\mathrm{acc},2}]\label{app:safsda}.
\end{align}

We will show the equality of \eqref{app:safsda1}.
Lemma \ref{app:VL0} implies
\begin{align}
G[\rho_{R1A1}, \Pi_{\mathrm{acc},1}]\ge 0,\quad
G[\rho_{R2A2}, \Pi_{\mathrm{acc},2}]\ge 0\label{app:SDJ}.
\end{align}
The combination of \eqref{app:SDJ} and \eqref{app:DJS} implies
\begin{align}
W_2 \ge 0 \label{app:BNS}.
\end{align}

The dual problem of \eqref{app:BNJ} is
\begin{align}
p_{\max}[\rho_{RA}, \Pi_{\mathrm{acc}}]
=
\min\{\Tr W |
W\otimes I_{B} - G[\rho_{RA}, \Pi_{\mathrm{acc}}]\ge 0
\}\label{app:safsda2}.
\end{align}
We denote the miximizers of \eqref{app:BNJ} with
$(\rho_{R1A1}, \Pi_{\mathrm{acc},1})$ and 
$(\rho_{R2A2}, \Pi_{\mathrm{acc},2})$ by
$C_1$ and $C_2$, respectively.
Similarly, we denote the minimizers of \eqref{app:safsda2} with
$(\rho_{R1A1}, \Pi_{\mathrm{acc},1})$ and 
$(\rho_{R2A2}, \Pi_{\mathrm{acc},2})$ by
$W_1$ and $W_2$, respectively.
Hence, we have
\begin{align}
W_1\otimes I_{B1} - G[\rho_{R1A1}, \Pi_{\mathrm{acc},1}]
&\ge 0 \label{app:DJS1}\\
W_2\otimes I_{B2} - G[\rho_{R2A2}, \Pi_{\mathrm{acc},2}]
&\ge 0 \label{app:DJS}\\
\Tr C_1 (W_1\otimes I_{B1} - G[\rho_{R1A1}, \Pi_{\mathrm{acc},1}])
&= 
\Tr W_1 -\Tr C_1 G[\rho_{R1A1}, \Pi_{\mathrm{acc},1}]=0
 \label{app:BN1} \\
\Tr C_2 (W_2\otimes I_{B2} - G[\rho_{R2A2}, \Pi_{\mathrm{acc},2}])
&= \Tr W_2 -\Tr C_2 G[\rho_{R2A2}, \Pi_{\mathrm{acc},2}]=0.\label{app:BN2}
\end{align}

Then, we have
\begin{align}
&W_1\otimes I_{B1}\otimes W_2\otimes I_{B2} - G[\rho_{R1A1}, \Pi_{\mathrm{acc},1}]\otimes G[\rho_{R2A2}, \Pi_{\mathrm{acc},2}]
\notag\\
=&
(W_1\otimes I_{B1} - G[\rho_{R1A1}, \Pi_{\mathrm{acc},1}])
\otimes 
W_2\otimes I_{B2}
+
G[\rho_{R1A1}, \Pi_{\mathrm{acc},1}]\otimes
(W_2\otimes I_{B2} - G[\rho_{R2A2}, \Pi_{\mathrm{acc},2}])\notag\\
\ge &0, \label{app:CSH}
\end{align}
where the final inequality follows from 
\eqref{app:DJS1},
\eqref{app:DJS},
\eqref{app:SDJ}, and
\eqref{app:BNS}.

Also, using \eqref{app:BN1} and \eqref{app:BN2}, we have
\begin{align}
&\Tr W_1\otimes W_2 - 
\Tr (C_1 \otimes C_2)
G[\rho_{R1A1}, \Pi_{\mathrm{acc},1}]\otimes G[\rho_{R2A2}, \Pi_{\mathrm{acc},2}]
\notag\\
=&
\Tr (C_1 \otimes C_2)(
W_1\otimes I_{B1}\otimes W_2\otimes I_{B2} - G[\rho_{R1A1}, \Pi_{\mathrm{acc},1}]\otimes G[\rho_{R2A2}, \Pi_{\mathrm{acc},2}]
)\notag\\
=&
\Tr (C_1 \otimes C_2)(
(W_1\otimes I_{B1} - G[\rho_{R1A1}, \Pi_{\mathrm{acc},1}])
\otimes 
W_2\otimes I_{B2}
+
G[\rho_{R1A1}, \Pi_{\mathrm{acc},1}]\otimes
(W_2\otimes I_{B2} - G[\rho_{R2A2}, \Pi_{\mathrm{acc},2}]))\notag\\
=&
\Tr (C_1 \otimes C_2)(
(W_1\otimes I_{B1} - G[\rho_{R1A1}, \Pi_{\mathrm{acc},1}])
\otimes 
W_2\otimes I_{B2})\notag\\
&+
\Tr (C_1 \otimes C_2)(
G[\rho_{R1A1}, \Pi_{\mathrm{acc},1}]\otimes
(W_2\otimes I_{B2} - G[\rho_{R2A2}, \Pi_{\mathrm{acc},2}]))\notag\\
=&
(\Tr C_1 (W_1\otimes I_{B1} - G[\rho_{R1A1}, \Pi_{\mathrm{acc},1}]))
\Tr C_2 (W_2\otimes I_{B2})\notag\\
&+
(\Tr C_1 
G[\rho_{R1A1}, \Pi_{\mathrm{acc},1}])
(\Tr C_2
(W_2\otimes I_{B2} - G[\rho_{R2A2}, \Pi_{\mathrm{acc},2}]))\notag\\
= &0.\label{app:ASD}
\end{align}
We have
\begin{align}
&p_{\max}[\rho_{R1A1}\otimes \rho_{R2A2}, 
\Pi_{\mathrm{acc},1}\otimes \Pi_{\mathrm{acc},2}] \notag\\
=&
\max_{C \ge 0}
\{\Tr_{A1A2B1B2} 
(G[\rho_{R1A1}, \Pi_{\mathrm{acc}1}] \otimes
G[\rho_{R2A2}, \Pi_{\mathrm{acc}2}] )
C
| \Tr_{B1B2} C=I_{A1A2} \} \notag\\
\stackrel{(a)}{\ge} &
\Tr (C_1 \otimes C_2)
G[\rho_{R1A1}, \Pi_{\mathrm{acc},1}]\otimes G[\rho_{R2A2}, \Pi_{\mathrm{acc},2}]\notag\\
\stackrel{(b)}{=}&\Tr W_1\otimes W_2 \notag\\
\stackrel{(c)}{\ge} &
\min\{\Tr W |
W\otimes I_{B1B2} - 
G[\rho_{R1A1}, \Pi_{\mathrm{acc}1}] \otimes
G[\rho_{R2A2}, \Pi_{\mathrm{acc}2}] 
\ge 0\}\notag\\
=&p_{\max}[\rho_{R1A1}\otimes \rho_{R2A2}, 
\Pi_{\mathrm{acc},1}\otimes \Pi_{\mathrm{acc},2}] ,
\end{align}
where $(a)$ follows from the relations
$C_1 \otimes C_2 \ge 0 $
and $\Tr_{B1B2} C_1 \otimes C_2=I_{A1A2}$.
$(b)$ follows from \eqref{app:ASD}.
$(c)$ follows from \eqref{app:CSH}.

Thus, we have
\begin{align}
&p_{\max}[\rho_{R1A1}\otimes \rho_{R2A2}, 
\Pi_{\mathrm{acc},1}\otimes \Pi_{\mathrm{acc},2}] \notag\\
=&
\Tr (C_1 \otimes C_2)
G[\rho_{R1A1}, \Pi_{\mathrm{acc},1}]\otimes G[\rho_{R2A2}, \Pi_{\mathrm{acc},2}]\notag\\
=&
\Tr C_1 
G[\rho_{R1A1}, \Pi_{\mathrm{acc},1}]
\Tr C_2
 G[\rho_{R2A2}, \Pi_{\mathrm{acc},2}]\notag\\
 =&p_{\max}[\rho_{R1A1}, \Pi_{\mathrm{acc},1}]
p_{\max}[\rho_{R2A2}, \Pi_{\mathrm{acc},2}]\label{app:NME}.
\end{align}

\subsection{Proof of Lemma \ref{app:VL0}}
The positive semi-definite matrices
$\rho_{RA}$ and $\Pi_{\mathrm{acc}}$
are written as
the sums of rank-one matrices as
$\sum_{k} p_k \rho_{RA|k}$
and $\sum_{k'} \Xi_{RB|k'}$, where
$\rho_{RA|k}$ are pure states
and $ \Xi_{RB|k'}$
are rank-one positive semi-definite matrices due to their spectral decompositions. 
Then, we have
\begin{align}
G[\rho_{RA}, \Pi_{\mathrm{acc}}]
=
\sum_{k}p_k
G[\rho_{RA|k}, \Pi_{\mathrm{acc}}]
=
\sum_{k}p_k\sum_{k'}
G[\rho_{RA|k}, \Xi_{RB|k'}].
\end{align}
Therefore,
it is sufficient to show Lemma \ref{app:VL0}
when $\rho_{RA}$ and $\Pi_{\mathrm{acc}}$ are rank-one matrix. 

We write 
$\rho_{RA}=
\sum_{i'j'} \sum_{ij} x_{ij} \overline{x}_{i'j'} |i\rangle \langle i'|_R 
|j\rangle \langle j'|_A$,
and
$\Pi_{\mathrm{acc}} =
\sum_{i'l'} \sum_{il} y_{il} \overline{y}_{i'l'} |i\rangle \langle i'|_R 
|l\rangle \langle l'|_B$.
Then, we have
\begin{align}
G[\rho_{RA}, \Pi_{\mathrm{acc}}]
=&
\Tr_R
\Big(\sum_{i'j'} \sum_{ij} x_{ij} \overline{x}_{i'j'} |i\rangle \langle i'|_R 
|j\rangle \langle j'|_A\otimes I_B \Big)^{T_A}
\Big(
\sum_{i'l'} \sum_{il} y_{il} \overline{y}_{i'l'} |i\rangle \langle i'|_R 
|l\rangle \langle l'|_B
\otimes I_A
\Big) \notag\\
=&
\Tr_R
\Big(\sum_{i'j'} \sum_{ij} x_{ij} \overline{x}_{i'j'} |i\rangle \langle i'|_R 
|j'\rangle \langle j|_A\otimes I_B \Big)
\Big(
\sum_{i'l'} \sum_{il} y_{il} \overline{y}_{i'l'} |i\rangle \langle i'|_R 
|l\rangle \langle l'|_B
\otimes I_A
\Big) \notag\\
=&
\sum_{j,l}\sum_{j',l'}
(\sum_i \overline{x}_{ij}y_{il})(\sum_{i'} x_{i'j'} \overline{y}_{i'l'})
|j\rangle \langle j'|_A 
|l\rangle \langle l'|_B \notag\\
=&
\sum_{j,l}\sum_{j',l'} m_{jl} \overline{m}_{j'l'}
|j\rangle \langle j'|_A 
|l\rangle \langle l'|_B \ge 0
\end{align}
where $m_{jl}:=
\sum_i \overline{x}_{ij}y_{il}$.
%Then, we define the operator $G[\rho_{RA}, \Pi_{\mathrm{acc}}]:= 
%\Tr_R (\rho_{RA}^{T_A}\otimes I_B)(\Pi_{\mathrm{acc}}\otimes I_A)$,

\subsection{Linear programming with general cone}\label{app:Ap2-conic}
For the proof of Lemma \ref{app:VL1}, we summarize 
linear programming with general cone \cite{alizadeh2003socp,luenberger2021linear}
as follows.
We consider conic linear programming  (EP) with a general cone.
Let ${\cal X}$ and ${\cal Z}$ be real vector spaces.
Let ${\cal P}\subset {\cal X}$ be a cone. 
Let $A$ be a linear function from ${\cal X}$ to ${\cal Z}$.
Given $c \in {\cal X}^*$ and $b \in {\cal Z}$, we consider the following minimization.

\begin{align}
EP:= \min_{x \in {\cal P}}\{ c(x) | 
Ax=b\}.
\end{align}
As the duality problem, we consider the following maximization.
\begin{align}
EP^*:= \max_{w \in {\cal Z}^*  } \{ w(b)|
-A^* w +c \in {\cal P}^*\} 
\end{align}
When the pair of 
$x \in {\cal P}$ and $w \in {\cal Z}^* $ satisfy the constraints, 
we have $c(x) -w(b)=(-A^* w +c)(x)\ge 0 $.
Hence, the inequality
\begin{align}
EP \ge EP^*
\end{align}
holds. In particular, when 
$x_0 \in {\cal P}$ and $w_0 \in {\cal Z}^* $ are solutions, we have
\begin{align}
EP =c(x_0) =w_0(b) = EP^*.
\end{align}

\section{Robustness to finite memory, operational imperfections, and detection loss (proofs)}
\subsection{Robustness to finite memory, operational noise, and detection loss (with proofs)}

\subsubsection*{Operational model (summary).}
We consider three implementation effects: (i) finite-coherence storage of the one-copy public key via a channel $\Lambda_{\mathrm{mem}}(\tau)$; (ii) imperfect operations/measurements modeled by $\tilde Z_j=\mathcal N_j\circ Z_j$ and a perturbed verification effect $\tilde M_1$; and (iii) detection loss producing an erasure symbol $\bot$ with probability $\lambda$ (abort-on-erasure vs.\ reject-on-erasure).
The resulting renormalization of $(\alpha,\beta)$ and the corresponding (approximate) zero-knowledge guarantee are stated and proved below (see Theorem~\ref{app:thm:robust-params}).

\subsubsection*{Interpretation (conservative worst-case bounds).}
Throughout this section we use diamond-norm and operator-norm budgets to obtain \emph{conservative} sufficient bounds that are uniform over all input states and downstream tests. In concrete platforms, tighter (state- and measurement-dependent) bounds can often be obtained from calibrated noise models; our goal here is to provide a robust, assumption-light envelope that suffices for security accounting.

The main effects are summarized in Table~\ref{app:non-ideal}.
\begin{table}[htbp]\centering
\begin{tabular}{l|l}
\hline
\textbf{Design knob} & \textbf{Effect}\\
\hline
Memory dwell $\tau$ & Additive shift $\pm\epsilon_{\mathrm{mem}}(\tau)$ to $(\alpha,\beta)$\\
Gate/POVM budgets & Additive shift $\pm(\epsilon_{\mathrm{ops}}+\delta_{\mathrm{ver}})$\\
Detection loss & Abort: conditional $(\alpha,\beta)$ unchanged; Reject: affine scaling by $(1-\lambda)$\\
\hline
\end{tabular}
\caption{Effects under non-ideal conditions. The definitions of $\epsilon_{\mathrm{mem}}(\tau)$,
$\pm(\epsilon_{\mathrm{ops}}$, $\delta_{\mathrm{ver}})$ and $\lambda$ are given in this section.}\label{app:non-ideal}
\end{table}

\subsubsection*{Overview (noise and loss robustness).}
We bound the deviation of acceptance probabilities under memory noise by $\tfrac12\|\Lambda_{\mathrm{mem}}(\tau)-\mathrm{id}\|_{\diamond}$ (Lemma~\ref{app:lem:memory}). Additional imperfect operations and POVM implementations are accounted for via a standard hybrid (telescoping) replacement argument, yielding an additive error budget (Lemma~\ref{app:lem:P2-accept}). We also distinguish abort-on-erasure versus reject-on-erasure conventions for detection loss (Lemmas~\ref{app:lem:P3-abort}--\ref{app:lem:P3-reject}). These components are collected into Theorem~\ref{app:thm:robust-params}.

\subsubsection*{Notation and generic facts (used repeatedly).}
For a binary verification test with effect $M$ (i.e., POVM $\{I-M,M\}$) on a state $\rho$, the acceptance probability is $p_{\mathrm{acc}}(\rho,M):=\Tr[M\rho]$. 
We use these standard facts:

\begin{itemize}
\item \textbf{F1 (Dual formulation of the diamond-norm).} For any Hermitian-preserving map $\mathcal{D}$ and any effect $0\le M\le I$,
\[
\sup_{\rho}\bigl|\Tr[M\,\mathcal{D}(\rho)]\bigr|\ \le\ \tfrac{1}{2}\|\mathcal{D}\|_\diamond.
\]
(See, e.g., Watrous, \emph{The Theory of Quantum Information}, Theorem 3.55~\cite{WatrousTQI}.)

\item \textbf{F2 (POVM perturbation).} 
If an intended effect $M$ is implemented as $\tilde M$ with $\|M-\tilde M\|_\infty\le \delta$, then for any state $\rho$,
\[
\bigl|\,\Tr[\tilde M\,\rho]-\Tr[M\,\rho]\,\bigr|\ \le\ \delta.
\]
(Immediate from Hölder inequality $\bigl|\Tr[A B]\bigr|\le \|A\|_\infty\|B\|_1$ with $\|\,\rho\,\|_1=1$.)

\item \textbf{F3 (Hybrid/triangle argument for composition).} If an intended channel $Z$ is implemented as $\tilde Z=\mathcal N\circ Z$ with $\tfrac{1}{2}\|\mathcal N-\mathrm{id}\|_\diamond\le \epsilon$, then for any test downstream of $\tilde Z$ the acceptance probability shifts by at most $\epsilon$. For a circuit with several noisy locations indexed by $j$, the total shift is at most $\sum_j \epsilon_j$ by a telescoping/hybrid argument (Watrous, \S3.3~\cite{WatrousTQI}).

\item \textbf{F4 (Chernoff--Hoeffding).} If $X_1,\dots,X_\kappa$ are i.i.d.\ Bernoulli with $\mathbb E[X_i]=p$, then for any $\theta\in(0,1)$,
\[
\Pr\!\left[\frac{1}{\kappa}\sum_{i=1}^\kappa X_i \le \theta\right]\ \le\ \exp\!\bigl(-2\kappa(\,p-\theta\,)^2\bigr),\qquad
\Pr\!\left[\frac{1}{\kappa}\sum_{i=1}^\kappa X_i \ge \theta\right]\ \le\ \exp\!\bigl(-2\kappa(\,\theta-p\,)^2\bigr).
\]
\end{itemize}

\subsubsection{Finite coherence time of quantum memories}
\subsubsection*{Finite-coherence memory (one-copy public key).}
In our one-copy setting, the verifier stores a single quantum public-key state $pk$ and later measures it during verification. A finite coherence time can be modeled by a storage channel $\Lambda_{\mathrm{mem}}(\tau)$ acting on $pk$ over a dwell time $\tau$, and we quantify its worst-case impact on any downstream accept/reject test by the diamond-norm budget $\epsilon_{\mathrm{mem}}(\tau)$. The next lemma formalizes the resulting additive perturbation bound.

\begin{lem}[Memory channel shifts acceptance by diamond budget]\label{app:lem:memory}
Let the verifier’s stored public key $pk$ (possibly entangled with an environment) undergo a memory channel $\Lambda_{\mathrm{mem}}(\tau)$. Let $M$ be the (ideal) verification effect applied after all ideal operations. Then
\begin{equation}\label{app:eq:P1-core}
\bigl|\,p_{\mathrm{acc}}^{\mathrm{noisy}}-p_{\mathrm{acc}}^{\mathrm{ideal}}\,\bigr|
=\bigl|\,\Tr\!\big[M\,(\Lambda_{\mathrm{mem}}(\tau)-\mathrm{id})(\rho)\big] \,\bigr|
\ \le\ \tfrac{1}{2}\,\bigl\|\Lambda_{\mathrm{mem}}(\tau)-\mathrm{id}\bigr\|_{\diamond}
\ :=:\ \epsilon_{\mathrm{mem}}(\tau).
\end{equation}
\end{lem}
\begin{proof}
Define the Hermitian-preserving map $\mathcal{D}:=\Lambda_{\mathrm{mem}}(\tau)-\mathrm{id}$. The verifier’s final acceptance probability is of the form $\Tr[M\,\sigma]$ for some post-memory state $\sigma$ on the verifier’s registers (possibly including a reference system purifying $pk$). By linearity,
\(p_{\mathrm{acc}}^{\mathrm{noisy}}-p_{\mathrm{acc}}^{\mathrm{ideal}}=\Tr[M\,\mathcal{D}(\rho)]\).
Applying Fact~F1 (diamond-norm duality) with the effect $M$ and the map $\mathcal{D}$ yields
\(\sup_{\rho}|\Tr[M\,\mathcal{D}(\rho)]|\le \tfrac12\|\mathcal{D}\|_{\diamond}\).
Since the supremum upper-bounds the particular state $\rho$ generated in the protocol, the bound~\eqref{app:eq:P1-core} follows.
\end{proof}

\begin{cor}[Parameter renormalization under finite memory]
If the ideal VIS/QZKP instance is $QZKP^t_{\alpha,\beta}$, then with memory dwell time $\tau$ it remains $QZKP^t_{\alpha-\epsilon_{\mathrm{mem}}(\tau),\ \beta+\epsilon_{\mathrm{mem}}(\tau)}$.
\end{cor}
\begin{proof}
For honest behavior, acceptance can decrease by at most $\epsilon_{\mathrm{mem}}(\tau)$, hence completeness is at least $\alpha-\epsilon_{\mathrm{mem}}(\tau)$. For any cheating strategy, the maximal achievable acceptance can increase by at most $\epsilon_{\mathrm{mem}}(\tau)$, hence soundness is at most $\beta+\epsilon_{\mathrm{mem}}(\tau)$.
\end{proof}
\subsubsection*{Examples.}
For dephasing with rate $\gamma$, $\epsilon_{\mathrm{mem}}(\tau)\le \tfrac12(1-e^{-\gamma\tau})$; for depolarizing with parameter $p(\tau)$, $\epsilon_{\mathrm{mem}}(\tau)\le p(\tau)$ (Nielsen--Chuang, \S 8.3.4~\cite{NielsenChuang}; Watrous, 
\S 4.1.2~\cite{WatrousTQI}).
\subsubsection*{Time-bounded refresh.}
Fix a budget $\varepsilon_{\mathrm{bud}}$ and restrict usage to $\tau^\star$ with $\epsilon_{\mathrm{mem}}(\tau^\star)\le \varepsilon_{\mathrm{bud}}$. Under $\kappa$ independent repetitions, a union/hybrid bound yields at most $\kappa\varepsilon_{\mathrm{bud}}$ total additive shift.
In this section, $\kappa$ denotes the number of thresholding trials.

\subsubsection{Gate/measurement imperfections: sufficient conditions and approximate ZK}
In this subsection we bound the impact of operational imperfections on acceptance probabilities and on the verifier’s view via a standard hybrid (telescoping) replacement argument. We then derive sufficient conditions for thresholded repetition (Chernoff--Hoeffding) and show that CSV/QSV zero-knowledge degrades gracefully to an $\epsilon_{\mathrm{zk}}$-approximate form measured in the verifier-view distance (Definition~\ref{app:def:view}).

\subsubsection*{Setup.}
Each intended map $Z_j$ is implemented as $\tilde Z_j=\mathcal N_j\circ Z_j$ with
\(\epsilon_j:=\tfrac12\|\mathcal N_j-\mathrm{id}\|_\diamond\).
Let $\epsilon_{\mathrm{ops}}:=\sum_j \epsilon_j$, where in our three-step VIS instance $j=1,2$ correspond to the verifier and prover operations prior to the final test.
The intended verification effect $M_1$ is implemented as $\tilde M_1$ with
\(\delta_{\mathrm{ver}}:=\|\tilde M_1-M_1\|_\infty\)
(note $0\le \delta_{\mathrm{ver}}\le 1$ since $0\le M_1,\tilde M_1\le I$).
If noise is correlated across locations, it can be absorbed into a single effective CPTP map on the corresponding joint system and bounded by one diamond-norm budget; the additive form is a conservative bookkeeping for a chosen decomposition.

\begin{lem}[Single repetition—acceptance perturbation]\label{app:lem:P2-accept}
For a single repetition of the base VIS protocol, letting $\Delta:=\epsilon_{\mathrm{mem}}+\epsilon_{\mathrm{ops}}+\delta_{\mathrm{ver}}$, we have
\[
\alpha_{\mathrm{eff}}\ \ge\ \alpha-\Delta,\qquad
\beta_{\mathrm{eff}}\ \le\ \beta+\Delta.
\]
\end{lem}
\begin{proof}
We formalize the “telescoping” argument by introducing a sequence of intermediate implementations.
Let $\Pi^{(0)}$ denote the ideal implementation of one repetition and $\Pi^{(K)}$ the fully noisy one. Construct a sequence
$\Pi^{(0)},\Pi^{(1)},\ldots,\Pi^{(K)}$
so that $\Pi^{(i)}$ and $\Pi^{(i+1)}$ differ in exactly one replacement: either (a) inserting the memory channel $\Lambda_{\mathrm{mem}}$, (b) replacing one $Z_j$ by $\tilde Z_j=\mathcal N_j\circ Z_j$, or (c) replacing the ideal effect $M_1$ by $\tilde M_1$.
Let $p_i$ be the verifier’s acceptance probability under $\Pi^{(i)}$. Then
\[
|p_K-p_0|\le \sum_{i=0}^{K-1}|p_{i+1}-p_i|\quad\text{(triangle inequality).}
\]
We bound each increment:

\emph{(a) Memory insertion.} By Lemma~\ref{app:lem:memory}, inserting $\Lambda_{\mathrm{mem}}$ shifts acceptance by at most $\epsilon_{\mathrm{mem}}$.

\emph{(b) Channel replacement.} Consider replacing $Z_j$ by $\tilde Z_j=\mathcal N_j\circ Z_j$. The rest of the protocol (including the final measurement) is a CPTP post-processing after this location. By Fact~F1 (diamond-norm duality) applied to the map $\mathcal{D}:=\mathcal N_j-\mathrm{id}$ and contractivity under CPTP post-processing, the acceptance probability changes by at most $\tfrac12\|\mathcal N_j-\mathrm{id}\|_\diamond=\epsilon_j$. Summing over all such locations yields the $\epsilon_{\mathrm{ops}}$ contribution.

\emph{(c) Effect replacement.} Replacing $M_1$ by $\tilde M_1$ changes acceptance by
\(|\Tr[(\tilde M_1-M_1)\rho]|\le \|\tilde M_1-M_1\|_\infty=\delta_{\mathrm{ver}}\)
for any state $\rho$ (Fact~F2 / Hölder inequality).

Combining the three contributions gives $|p_K-p_0|\le \epsilon_{\mathrm{mem}}+\epsilon_{\mathrm{ops}}+\delta_{\mathrm{ver}}=\Delta$. Applying this bound to the honest strategy (lower deviation) and to the optimal cheating strategy (upper deviation) yields the stated inequalities for $(\alpha_{\mathrm{eff}},\beta_{\mathrm{eff}})$.
\end{proof}

\subsubsection*{Repetition with thresholding.}
Let $X_i\in\{0,1\}$ be the indicator for acceptance in repetition $i$, and accept iff $\sum_i X_i\ge \tau_{\mathrm{acc}}$.
The number of repetitions is denoted by $\kappa$.
Under i.i.d. modeling with
\[
 p_{\mathrm{H}}\ \ge\ \alpha-\Delta,\qquad p_{\mathrm{F}}\ \le\ \beta+\Delta,
\]
Chernoff--Hoeffding (F4) yields
\[
\Pr[\text{honest reject}] \le \exp\!\bigl(-2\kappa(\,p_{\mathrm{H}}-\tfrac{\tau_{\mathrm{acc}}}{\kappa}\,)^2\bigr),\quad
\Pr[\text{forger accept}] \le \exp\!\bigl(-2\kappa(\,\tfrac{\tau_{\mathrm{acc}}}{\kappa}-p_{\mathrm{F}}\,)^2\bigr).
\]
A sufficient design condition is $p_{\mathrm{F}}<\tfrac{\tau_{\mathrm{acc}}}{\kappa}<p_{\mathrm{H}}$. Choosing
\(\tau_{\mathrm{acc}}=\left\lceil \kappa\,\tfrac{p_{\mathrm{H}}+p_{\mathrm{F}}}{2}\right\rceil\)
(mid-gap thresholding) equalizes the two exponents.

\begin{defn}[Verifier view]\label{app:def:view}
Fix one repetition of the VIS/QZKP interaction. Let $\mathcal{T}$ denote all
\emph{classical} transcript registers that are available to the verifier
(including any classical challenges it sent, classical responses it received, and
classical measurement outcomes it registered), and let $\mathcal{W}$ denote the
verifier’s private \emph{quantum} workspace just before producing the final
accept/reject bit. The \emph{view} of the verifier is the (classical–quantum)
state
\[
\mathsf{View} \;:=\; \rho_{\mathcal{T}\mathcal{W}}
\]
obtained by tracing out all registers not held by the verifier (i.e., prover
systems, discarded environments, etc.). When the transcript is purely classical
we identify $\mathsf{View}$ with the induced probability distribution on
$\mathcal{T}$. The statistical distance $\mathrm{SD}(\cdot,\cdot)$ denotes total
variation distance in the classical case, and, for cq‑states, the standard
quantum generalization via the trace distance:
\[
\mathrm{SD}(\rho,\sigma) \;:=\; \tfrac12\|\rho-\sigma\|_1.
\]
(Statistical distance is contractive (non-increasing) under CPTP maps and classical post‑processing~\cite{WatrousTQI}.)
\end{defn}

This notion concerns the final verifier view for one repetition (immediately before output); a step-wise variant can be defined analogously if needed.

\begin{lem}[Approximate CSV/QSV zero-knowledge under small operational noise]\label{app:lem:P2-ZK}
Let $\mathsf{View}_{\mathrm{noisy}}$ and $\mathsf{View}_{\mathrm{ideal}}$ be the verifier views for one repetition of the noisy and ideal protocols. Suppose the implementation budgets satisfy
\[
\epsilon_{\mathrm{mem}}:=\tfrac12\|\Lambda_{\mathrm{mem}}-\mathrm{id}\|_\diamond,\quad
\epsilon_{\mathrm{ops}}:=\sum_j \tfrac12\|\mathcal{N}_j-\mathrm{id}\|_\diamond,\quad
\delta_{\mathrm{ver}}:=\|\tilde M_1-M_1\|_\infty,
\]
and set $\epsilon_{\mathrm{zk}}:=\epsilon_{\mathrm{mem}}+\epsilon_{\mathrm{ops}}+\delta_{\mathrm{ver}}$.
If the ideal protocol is CSV/QSV zero knowledge with simulator $\mathsf{Sim}$ (i.e., $\mathrm{SD}(\mathsf{View}_{\mathrm{ideal}},\mathsf{Sim})=0$), then the noisy protocol is $\epsilon_{\mathrm{zk}}$-approximate CSV/QSV zero knowledge:
\[
\mathrm{SD}(\mathsf{View}_{\mathrm{noisy}},\mathsf{Sim})\le \epsilon_{\mathrm{zk}}.
\]
For $\kappa$ independent repetitions combined by a fixed deterministic threshold post-processing,
$\mathrm{SD}(\mathsf{View}^{(\kappa)}_{\mathrm{noisy}},\mathsf{Sim}^{(\kappa)})\le O(\kappa)\,\epsilon_{\mathrm{zk}}$.
\end{lem}

\noindent\emph{Clarification.}
Here zero-knowledge is measured in the verifier-view distance $\mathrm{SD}$ of Definition~\ref{app:def:view}; the simulator $\mathsf{Sim}$ is the same ideal-world simulator, and implementation imperfections only contribute an additional trace-distance deviation between $\mathsf{View}_{\mathrm{noisy}}$ and $\mathsf{View}_{\mathrm{ideal}}$.

\begin{proof}
We again use a hybrid argument. Construct intermediate implementations $\Pi^{(0)},\ldots,\Pi^{(K)}$ from ideal to fully noisy, differing by exactly one replacement each time (memory insertion, one $Z_j$ replacement, or the $M_1$ effect replacement). Let $\mathsf{View}^{(i)}$ be the verifier view induced by $\Pi^{(i)}$.
By the triangle inequality for trace distance,
\[
\mathrm{SD}(\mathsf{View}_{\mathrm{noisy}},\mathsf{View}_{\mathrm{ideal}})
=\mathrm{SD}(\mathsf{View}^{(K)},\mathsf{View}^{(0)})
\le \sum_{i=0}^{K-1}\mathrm{SD}(\mathsf{View}^{(i+1)},\mathsf{View}^{(i)}).
\]
We bound each increment.

\emph{Memory insertion:} inserting $\Lambda_{\mathrm{mem}}$ changes the view by at most $\epsilon_{\mathrm{mem}}$ by Fact~F1 applied to the map $\mathcal{D}:=\Lambda_{\mathrm{mem}}-\mathrm{id}$ and contractivity of trace distance under CPTP post-processing.

\emph{Channel replacements:} replacing one $Z_j$ by $\tilde Z_j=\mathcal N_j\circ Z_j$ changes the view by at most $\tfrac12\|\mathcal N_j-\mathrm{id}\|_\diamond$. Summing over $j$ yields $\epsilon_{\mathrm{ops}}$.

\emph{Effect replacement:} replacing $M_1$ by $\tilde M_1$ changes the statistics of any classical transcript bit produced by this measurement by at most $\|\tilde M_1-M_1\|_\infty=\delta_{\mathrm{ver}}$ (Fact~F2). Since the view is a cq-state whose classical part is obtained by measuring with this effect and whose quantum workspace is obtained by a CPTP map conditioned on the transcript, the overall trace distance between the cq-views is also bounded by $\delta_{\mathrm{ver}}$ (data processing / contractivity under conditioned CPTP maps).

Therefore, summing the three contributions gives
\(\mathrm{SD}(\mathsf{View}_{\mathrm{noisy}},\mathsf{View}_{\mathrm{ideal}})\le \epsilon_{\mathrm{mem}}+\epsilon_{\mathrm{ops}}+\delta_{\mathrm{ver}}=\epsilon_{\mathrm{zk}}\).
Finally, by triangle inequality,
\(
\mathrm{SD}(\mathsf{View}_{\mathrm{noisy}},\mathsf{Sim})
\le \mathrm{SD}(\mathsf{View}_{\mathrm{noisy}},\mathsf{View}_{\mathrm{ideal}})
+\mathrm{SD}(\mathsf{View}_{\mathrm{ideal}},\mathsf{Sim})
\le \epsilon_{\mathrm{zk}},
\)
using $\mathrm{SD}(\mathsf{View}_{\mathrm{ideal}},\mathsf{Sim})=0$.

For $\kappa$ independent repetitions, we apply the same telescoping trick on tensor products:
\(\mathrm{SD}(\rho_1\otimes\cdots\otimes\rho_\kappa,\sigma_1\otimes\cdots\otimes\sigma_\kappa)\le \sum_{t=1}^\kappa \mathrm{SD}(\rho_t,\sigma_t)\), hence the distance scales at most linearly in $\kappa$. Any fixed deterministic thresholding is a classical post-processing map, and trace distance is contractive under CPTP maps, so the bound persists after thresholding.
\end{proof}

\subsubsection{Detection efficiency / loss modeling (abort vs.\ reject)}
We model detection loss at the decisive verification measurement as an erasure outcome $\bot$ with probability $\lambda$, and distinguish two conventions: abort-on-erasure (discard and optionally retry) versus reject-on-erasure (map $\bot$ to outcome~0). We show how each convention induces effective parameters, separating the effect on conditional security from the effect on throughput.
Let $\lambda$ denote the probability that the decisive verification measurement yields an erasure symbol $\bot$ (no click).

We treat $\lambda$ as a device parameter that upper-bounds the no-click probability uniformly, i.e., $\Pr[\bot]\le \lambda$ for all input states (and strategies) at the decisive detection.

\subsubsection*{Modeling assumption (erasure flag).}
Throughout this subsection, $\bot$ is treated as an explicit \emph{erasure flag}: with probability $\lambda$ the measurement produces $\bot$, and conditional on a click (i.e., no erasure) the accept/reject bit is generated by the same (noisy) verification POVM effect $\tilde M_1$ as in Lemma~\ref{app:lem:P2-accept}. In particular, under this erasure-flag model the erasure event is independent of which $\{0,1\}$ outcome would be produced upon a click. If one does not wish to assume erasure--outcome independence (e.g.\ due to thresholding-induced correlations), one may adopt the reject-on-erasure convention (Lemma~\ref{app:lem:P3-reject}) to obtain conservative unconditional bounds.

\subsubsection*{Abort-on-erasure.}
In this convention, we post-select on completed runs (no erasure) and optionally retry otherwise. The next lemma formalizes that, under the erasure-flag assumption above, post-selection affects only throughput, while the conditional acceptance parameters follow Lemma~\ref{app:lem:P2-accept}.
\begin{lem}[Abort-on-erasure]\label{app:lem:P3-abort}
If runs with outcome $\bot$ are discarded and may be retried, then the \emph{conditional} acceptance parameters $(\alpha,\beta)$ for completed runs remain those of Lemma~\ref{app:lem:P2-accept}:
\[
\alpha_{\mathrm{cond}}\ge \alpha-\Delta,\qquad \beta_{\mathrm{cond}}\le \beta+\Delta.
\]
Unconditional throughput is multiplied by the survival factor $(1-\lambda)$ per affected step.
\end{lem}
\begin{proof}
Under abort-on-erasure, we condition on the event $E$ that the decisive measurement produced a click (i.e., did not output $\bot$). By definition of this convention, all runs with $E^c$ are discarded; security statements are made for the conditional distribution given $E$.
The physical implementation (memory+operations+verification) for a completed run is exactly the same as in Lemma~\ref{app:lem:P2-accept}, except that we restrict to the subensemble where a decision is produced. Therefore the same additive perturbation bound applies to the conditional acceptance probabilities, yielding the stated conditional parameters.
The only effect of loss is that a run completes with probability $\Pr[E]=1-\lambda$, so the throughput (rate of completed decisions) is scaled by $(1-\lambda)$.
\end{proof}

\subsubsection*{Reject-on-erasure.}
This convention provides \emph{unconditional} accounting without post-selection by mapping $\bot$ to reject. Consequently, both completeness and soundness are scaled by the click probability $(1-\lambda)$, as captured below.
\begin{lem}[Reject-on-erasure]\label{app:lem:P3-reject}
If $\bot$ is mapped to outcome $0$ (reject), then for a single repetition
\[
\alpha_{\mathrm{rej}} \ge (1-\lambda)\,(\alpha-\Delta),\qquad
\beta_{\mathrm{rej}} \le (1-\lambda)\,(\beta+\Delta).
\]
\end{lem}
\begin{proof}
Under reject-on-erasure, acceptance requires both (i) a click and (ii) acceptance conditioned on a click. Thus for honest behavior,
\[
 p^{(\mathrm{H})}_{\mathrm{acc,rej}}=\Pr[\text{click}]\,\Pr[\text{accept}\mid \text{click}]
\ge (1-\lambda)\,(\alpha-\Delta),
\]
where we used $\Pr[\text{click}]=1-\lambda$ and Lemma~\ref{app:lem:P2-accept} for the conditional acceptance.
Similarly, for any cheating strategy,
\[
 p^{(\mathrm{F})}_{\mathrm{acc,rej}}\le \Pr[\text{click}]\,\Pr[\text{accept}\mid \text{click}]
\le (1-\lambda)\,(\beta+\Delta),
\]
again using Lemma~\ref{app:lem:P2-accept} for the conditional acceptance bound.
\end{proof}

\subsubsection*{Repetition with thresholding under loss.}
Under abort-on-erasure, the Chernoff analysis from the previous subsection applies to the conditional (completed-run) Bernoulli trials, while throughput is reduced by $(1-\lambda)$.
Under reject-on-erasure, replace the Bernoulli parameters by
\[
 p_{\mathrm{H}}\gets (1-\lambda)(\alpha-\Delta),\qquad
 p_{\mathrm{F}}\gets (1-\lambda)(\beta+\Delta),
\]
and choose $\kappa,\tau_{\mathrm{acc}}$ to maintain $p_{\mathrm{F}}<\tfrac{\tau_{\mathrm{acc}}}{\kappa}<p_{\mathrm{H}}$.

\subsubsection{Collected theorem (proved)}
We now collect the preceding bounds into a single statement that summarizes the effective completeness/soundness parameters under memory noise, operational imperfections, and detection loss, together with the corresponding concentration bounds under thresholded repetition and the resulting approximate zero-knowledge guarantee.
\begin{thm}[Noisy-and-lossy parameters for VIQDS from VIS]\label{app:thm:robust-params}
Let the ideal VIS/QZKP instance have $(\alpha,\beta)$. Suppose the verifier’s one-copy public key experiences memory noise with budget $\epsilon_{\mathrm{mem}}(\tau)$, the operations incur $\epsilon_{\mathrm{ops}}$ and the verification effect budget is $\delta_{\mathrm{ver}}:=\|\tilde M_1-M_1\|_\infty$, and the decisive verification has erasure probability bounded by $\lambda$. For a single repetition set
\[\Delta:=\epsilon_{\mathrm{mem}}+\epsilon_{\mathrm{ops}}+\delta_{\mathrm{ver}}.\]
Then:
\begin{enumerate}
\item \textbf{No erasure or abort-on-erasure:} $\alpha_{\mathrm{eff}}\ge \alpha-\Delta$, $\beta_{\mathrm{eff}}\le \beta+\Delta$.
\item \textbf{Reject-on-erasure:} $\alpha_{\mathrm{rej}}\ge (1-\lambda)(\alpha-\Delta)$ and $\beta_{\mathrm{rej}}\le (1-\lambda)(\beta+\Delta)$.
\item \textbf{Repetition with threshold $\tau_{\mathrm{acc}}$.} If
\[
\beta' < \tfrac{\tau_{\mathrm{acc}}}{\kappa} < \alpha'\quad
\text{for}\quad
(\alpha',\beta')=\begin{cases}
(\alpha-\Delta,\ \beta+\Delta) & \text{abort-on-erasure},\\
((1-\lambda)(\alpha-\Delta),\ (1-\lambda)(\beta+\Delta)) & \text{reject-on-erasure},
\end{cases}
\]
then
\[
\Pr[\text{honest reject}] \le e^{-2\kappa(\alpha'-\tfrac{\tau_{\mathrm{acc}}}{\kappa})^2},\qquad
\Pr[\text{forger accept}] \le e^{-2\kappa(\tfrac{\tau_{\mathrm{acc}}}{\kappa}-\beta')^2}.
\]
\item \textbf{Approximate CSV/QSV zero-knowledge.} The one-shot simulation error is at most $\epsilon_{\mathrm{zk}}\le \Delta$ and grows as $O(\kappa)\Delta$ under $\kappa$ i.i.d. repetitions with fixed thresholding.
\end{enumerate}
\end{thm}
\begin{proof}
Item (1) is Lemma~\ref{app:lem:P2-accept} together with Lemma~\ref{app:lem:P3-abort} (conditioning on completion does not alter the conditional perturbation bound).
Item (2) is Lemma~\ref{app:lem:P3-reject}.
Item (3) applies Chernoff--Hoeffding (F4) to Bernoulli trials with success parameters bounded as in items (1) or (2).
Item (4) is Lemma~\ref{app:lem:P2-ZK}, and the $O(\kappa)$ scaling follows from subadditivity of trace distance on product states and contractivity under deterministic post-processing.
\end{proof}
\subsubsection*{Remark on independence.}
The Chernoff/Hoeffding exponents assume i.i.d.\ trials. This is naturally met when each repetition uses independently prepared tokens (fresh one-copy public-key instances and independent randomness) and the verifier applies a fixed threshold rule. If weak correlations exist across repetitions (e.g.\ slow drift), Azuma--Hoeffding or Bernstein-type bounds can be invoked under standard martingale/mixing assumptions; the essential takeaway (exponential decay in $\kappa$ as long as a gap remains) continues to hold under mild mixing assumptions.
In particular, in the one-copy setting, $\kappa$ repetitions require $\kappa$ independently prepared public-key tokens (and independent randomness), since each token is consumed by the verification measurement.

\subsection{Sufficient feasibility thresholds under non-idealities}
We provide \emph{explicit sufficient (conservative)} parameter regimes that keep the protocol within the robustness envelope of Theorem~\ref{app:thm:robust-params}. These closed-form inequalities are intended for budgeting and feasibility estimates, and can typically be improved under platform-specific calibrated noise models.

\subsubsection*{Notation recap.} For one repetition, let the total non-ideality be
$\Delta:=\epsilon_{\mathrm{mem}}+\epsilon_{\mathrm{ops}}+\delta_{\mathrm{ver}}$, where
$\epsilon_{\mathrm{mem}}=\tfrac12\|\Lambda_{\mathrm{mem}}-\mathrm{id}\|_\diamond$ is the storage
budget for the (one-copy) public key, $\epsilon_{\mathrm{ops}}=\sum_j \tfrac12\|\mathcal N_j-\mathrm{id}\|_\diamond$
aggregates the noisy implementations of the intended operations ($\sfZ(t)$, basis changes, etc.), and
$\delta_{\mathrm{ver}}=\|\tilde M_1-M_1\|_\infty$ is the verification-effect budget. The
ideal completeness and soundness are $(\alpha,\beta)=(1,1/d)$ with $d=p^l$ in our construction.
Let $\kappa$ be the number of repetitions.
Let $\lambda$ be the no-click probability (detection loss) at the decisive verification measurement;
$\eta:=1-\lambda$ is the detection efficiency.

\subsubsection*{Security gap and what must be preserved.} Under abort-on-erasure, the effective parameters are
$\alpha'\ge 1-\Delta$ and $\beta'\le 1/d+\Delta$. Define the 
\emph{gap} $G:=\alpha'-\beta'$, so
\begin{equation}\label{app:eq:gap-abort}
G_{\mathrm{abort}}\ \ge\ 1-\frac{1}{d}-2\Delta.
\end{equation}
Under reject-on-erasure, $\alpha'\ge (1-\lambda)(1-\Delta)$ and
$\beta'\le (1-\lambda)(1/d+\Delta)$, which yields
\begin{equation}\label{app:eq:gap-reject}
G_{\mathrm{reject}}\ \ge\ (1-\lambda)\Bigl(1-\frac{1}{d}-2\Delta\Bigr).
\end{equation}
A \emph{sufficient} condition to maintain the theoretical scaling $\epsilon\asymp 1/d$ (and to meet a
final target \emph{no larger than} $1/d$ after repetition/thresholding) is that the single-repetition
gap is strictly positive, $G>0$, so that standard Chernoff amplification drives the forging probability
below any target (including $1/d$) with overhead logarithmic in the target.

\subsubsection*{Minimum coherence time (time-bounded memory).}
Choose a design repetition level $\kappa$ (number of thresholding trials) and a target 
\emph{gap margin} $G_{\min}>0$.
Under abort-on-erasure, enforce
\begin{equation}\label{app:eq:Delta-budget}
\Delta\ \le\ \frac{1}{2}\Bigl(1-\frac{1}{d}-G_{\min}\Bigr).
\end{equation}
With a dephasing memory channel of rate $\gamma$ (i.e., $\epsilon_{\mathrm{mem}}(\tau)\le \tfrac12(1-e^{-\gamma\tau})$),
any dwell time $\tau\le \tau^{\star}$ with
\begin{equation}\label{app:eq:tau-star}
\tau^{\star}\ \le\ \frac{1}{\gamma}\,\ln\!\left(\frac{1}{1-2\,\epsilon_{\mathrm{mem}}^{\max}}\right),\quad\text{where }\ 
\epsilon_{\mathrm{mem}}^{\max}:=\frac{1}{2}\Bigl(1-\frac{1}{d}-G_{\min}\Bigr)-\epsilon_{\mathrm{ops}}-\delta_{\mathrm{ver}},
\end{equation}
meets~\eqref{app:eq:Delta-budget}. For a depolarizing memory with parameter $p(\tau)$, it suffices that
$p(\tau)\le \epsilon_{\mathrm{mem}}^{\max}$.

\subsubsection*{Acceptable gate/measurement error for discrete Heisenberg operations.}
A conservative sufficient budget is
\begin{equation}\label{app:eq:gate-budget}
\epsilon_{\mathrm{ops}}\le 2\,\epsilon_{g}^{\max},\qquad \delta_{\mathrm{ver}}\le \delta_{\mathrm{ver}}^{\max},\qquad
\epsilon_{\mathrm{mem}}\le \epsilon_{\mathrm{mem}}^{\max},\ 
\end{equation}
with \eqref{app:eq:Delta-budget} enforcing
\begin{equation}\label{app:eq:epsg-max}
\epsilon_{g}^{\max}\ \le\ \frac{\tfrac{1}{2}\bigl(1-\tfrac{1}{d}-G_{\min}\bigr)-\epsilon_{\mathrm{mem}}-\delta_{\mathrm{ver}}}{2}.
\end{equation}
Equation~\eqref{app:eq:epsg-max} directly converts the desired
security margin $G_{\min}$ into a per-gate error tolerance.

\subsubsection*{Minimum detection efficiency.}
Under reject-on-erasure, if we require a minimum effective completeness $\alpha'_{\min}$, then
\[
\alpha'_{\min} \le (1-\lambda)(1-\Delta)=\eta(1-\Delta),
\]
so a sufficient minimum detection efficiency is
\[
\eta_{\min}=\frac{\alpha'_{\min}}{1-\Delta}.
\]
Moreover, if we also impose a target maximum soundness $\beta'_{\max}$, then it suffices that
\[
\eta\Bigl(\frac{1}{d}+\Delta\Bigr)\le \beta'_{\max},
\]
i.e., $\eta \le \beta'_{\max}/(\frac{1}{d}+\Delta)$.

\subsubsection*{Hitting the target $\epsilon_{\mathrm{final}}\le 1/d$ with Chernoff amplification.}
Once $G\ge G_{\min}>0$ holds (abort or reject model with the corresponding $\Delta$, $\lambda$), choose a
threshold rule and total repetition count $\kappa$ so that
\begin{equation}\label{app:eq:ell-choice}
\Pr[\text{forger accept}]\ \le\ \exp\!\Bigl(-\tfrac{\kappa}{2}\,G_{\min}^2\Bigr)\ \le\ \frac{1}{d}.
\end{equation}
A sufficient choice is
$\kappa\ge \bigl(2\ln d\bigr)/G_{\min}^2$ with mid-gap thresholding (see Methods).

\subsubsection*{Summary of practical minima (sufficient conditions).}
Given a chosen number of repetition $\kappa$ and target single-shot gap $G_{\min}>0$:
\begin{enumerate}
  \item \textbf{Memory coherence (one-copy $pk$):} pick a time bound $\tau^{\star}$ such that $\epsilon_{\mathrm{mem}}(\tau^{\star})\le \epsilon_{\mathrm{mem}}^{\max}$ with $\epsilon_{\mathrm{mem}}^{\max}$ from~\eqref{app:eq:tau-star}.
  \item \textbf{Gate/measurement error per primitive:} ensure $\epsilon_g\le \epsilon_g^{\max}$ from~\eqref{app:eq:epsg-max} given $m$ and the chosen split of the total $\Delta$ budget.
  \item \textbf{Detection efficiency:} 
    \begin{itemize}
      \item Abort-on-erasure: no minimum efficiency is needed for security (only for throughput);
\item Reject-on-erasure: enforce $\eta \ge \eta_{\min}=\alpha'_{\min}/(1-\Delta)$ as stated in the paragraph “Minimum detection efficiency”.
    \end{itemize}
  \item \textbf{Repetition/thresholding:} pick $\kappa\ge (2\ln d)/G_{\min}^2$ to guarantee $\epsilon_{\mathrm{final}}\le 1/d$.
\end{enumerate}

These conditions are \emph{sufficient} (often conservative). In practice, tighter budgets can be extracted by replacing worst-case diamond-norm sums with calibrated, platform-specific error models and by using tighter concentration bounds or exact binomial tail computations for the chosen threshold rule.

\section{Quantitative comparison with existing QDS schemes}
This section provides an \emph{indicative} (order-of-magnitude) quantitative comparison
with representative quantum digital signature (QDS) demonstrations, intended to
contextualize \emph{signing rates} and \emph{classical signature lengths} under a single,
explicit accounting convention. Because existing QDS protocols differ substantially in
workflow (e.g., offline pre-distribution, post-processing, and what is counted as “one
signature”), the numbers below should be read as guidance rather than a fully standardized
benchmark. \cite{roberts2017,richter2021,yin2023,du2025,wang2026}

\subsubsection*{Accounting convention.}
Throughout, $O$ denotes the \emph{ideal} total number of \emph{transmitted} quantum systems
(qubits in our qubit-based instantiation) per signed message \emph{under our accounting}
(i.e., as determined by the protocol description and our counting rule).
$L_S$ denotes the total number of \emph{classical} signature bits transmitted per signed message,
$\epsilon$ denotes the reported failure/forgery upper bound, and $R_S$ denotes the signing rate.
Unless explicitly stated otherwise, we use values as reported in the cited experimental papers.
We do \emph{not} attempt to convert prior experiments into our accounting for $O$, since protocol
workflows and what is counted as “one signature” vary across implementations\cite{roberts2017,richter2021,yin2023,du2025,wang2026}.

\subsubsection*{Our scheme (qubit-based instantiation and message unit).}
To match near-term experimental capabilities, we consider the qubit-based instantiation obtained
by concatenating $l$ independent qubit-based QZKP instances (setting $p=2$ in the base protocol),
which yields soundness $\beta = 2^{-l}$.
A signed $L_{\mathrm{msg}}$-bit document is produced by repeating the 1-bit signing primitive
$L_{\mathrm{msg}}$ times (repeated single-bit scheme), and all resource counts below are totals
\emph{per $L_{\mathrm{msg}}$-bit document}. Under this accounting,
the total public-key length consumed per $L_{\mathrm{msg}}$-bit document is
$4lL_{\mathrm{msg}}$ qubits, and the total qubit transmission overhead per signed document is
\begin{equation}
O = 6lL_{\mathrm{msg}}.
\end{equation}
The total classical signature length transmitted per signed document is
\begin{equation}
L_S = lL_{\mathrm{msg}}.
\end{equation}

\subsubsection*{Indicative signing rate under a common optical model (for our scheme).}
Ignoring single-qubit gate times and classical communication time, we estimate an indicative
signing rate for our scheme as
\begin{equation}
R_S(D)=\frac{C\,\eta_d}{O}\times 10^{-\frac{\alpha D}{10}},
\end{equation}
where $C$ is the clock rate, $\eta_d$ is the detection efficiency, and $\alpha$ is the fiber attenuation
(in dB/km) at distance $D$ km.
(These platform parameters can be replaced by system-specific values without changing the scaling.)
We emphasize that this is a back-of-the-envelope throughput estimate under a \emph{common} optical
loss model, used only to contextualize our scheme; prior experiments are listed using their reported
figures without re-normalization across heterogeneous workflows. \cite{wang2026,roberts2017,richter2021,yin2023,du2025}

\noindent\textbf{Important accounting note.}
Because protocol workflows and what is counted as “one signature” differ across experiments,
Table~\ref{app:Comparison} should be interpreted as an indicative comparison under the stated
accounting convention, not as a fully normalized benchmark \cite{roberts2017,richter2021,yin2023,du2025,wang2026}.

\begin{table}[t]
\caption{Comparison with representative QDS experiments (indicative).}
\label{app:Comparison}
\centering
\scriptsize
\setlength{\tabcolsep}{6pt}
\renewcommand{\arraystretch}{1.05}
\begin{center}
\begin{tabular}{l l l l l l l l}
\hline
References & Protocol & Clock rate & Distance & Document size & $L_S$ & $\epsilon$ & $R_S$ \\
\hline
Roberts et al.~\cite{roberts2017} & MDI & 1 GHz & 50 km & 1 bit &
$2.5\times 10^{6}$ & $10^{-10}$ & $2.22\times 10^{-2}$ tps \\
\hline
Richter et al.~\cite{richter2021} & CV & 1 GHz & 20 km & 1 bit &
$2.08\times 10^{8}$ & $10^{-4}$ & 5 tps \\
\hline
Yin et al.~\cite{yin2023} & BB84 & 200 MHz & 101 km & $10^{6}$ bits &
256 & $10^{-32}$ & 1.22 tps \\
\hline
\multirow[t]{2}{*}{Du et al.~\cite{du2025}} & \multirow[t]{2}{*}{BB84} & \multirow[t]{2}{*}{50 MHz}
& 100 km & \multirow[t]{2}{*}{$10^{6}$ bits} &
783 & $4.64\times 10^{-8}$ & 6.50 tps \\
\cline{4-8}
& & & 200 km & &
1029 & $4.72\times 10^{-8}$ & $4.14\times 10^{-2}$ tps \\
\hline
Wang et al.~\cite{wang2026} & Single-qubit & 50 MHz & 100 km & $10^{6}$ bits &
$6\times 10^{8}$ & $3.7\times 10^{-3}$ & $6.9\times 10^{-5}$ tps \\
\hline
Our work & Our protocol & 50 MHz & 100 km & $10^{6}$ bits &
$3\times 10^{7}$ & $2^{-30}$ & $1.39\times 10^{-3}$ tps \\
\hline
\end{tabular}
\end{center}
\textit{$L_S$ represents the total classical signature bits per signed message,
$\epsilon$ represents the reported failure/forgery upper bound, and $R_S$ represents the signing rate.
For our scheme, the ideal transmitted-qubit overhead under our accounting is $O=6lL_{\mathrm{msg}}$
and is used only in the throughput estimate above (not re-normalized across prior experiments).}
\end{table}

\bibliographystyle{unsrt}
\bibliography{references}
\end{document}